\documentclass[aps,prd,twocolumn,superscriptaddress,preprintnumbers,nofootinbib,longbibliography,floatfix]{revtex4-1}

\pdfoutput=1

\usepackage[utf8]{inputenc}
\usepackage{graphicx}

\usepackage[colorlinks=true
,urlcolor=blue
,anchorcolor=blue
,citecolor=blue
,filecolor=blue
,linkcolor=blue
,menucolor=blue
,pagecolor=blue
]{hyperref}
\usepackage[all]{hypcap}

\newcommand{\xys}[1]{\textbf{\textcolor{darkred}{{***}}}}
\newcommand{\xms}[1]{\textbf{\textcolor{darkblue}{{***}}}}

\usepackage{amsmath}
\usepackage{amssymb}

\usepackage{multirow}

\newcommand{\nameofsearch}{$p_T$-enhanced }
\newcommand{\ptmm}{p_T^{\mu\mu}}

\newcommand{\GeV}{\text{~GeV}}

\newcommand{\BR}{\text{BR}}

\newcommand{\cL}{\mathcal{L}}
\newcommand{\inv}{$^{-1}$}

\DeclareRobustCommand{\Sec}[1]{Sec.~\ref{#1}}

\DeclareRobustCommand{\Tab}[1]{Table~\ref{#1}}

\DeclareRobustCommand{\Fig}[1]{Fig.~\ref{#1}}
\DeclareRobustCommand{\Figs}[2]{Figs.~\ref{#1} and \ref{#2}}
\DeclareRobustCommand{\Eq}[1]{Eq.~(\ref{#1})}

\DeclareRobustCommand{\Ref}[1]{Ref.~\cite{#1}}
\DeclareRobustCommand{\Refs}[1]{Refs.~\cite{#1}}

\newcommand{\be}{\begin{equation}}
\newcommand{\ee}{\end{equation}}

\begin{document}

\title{Searching in CMS Open Data for Dimuon Resonances \\ with Substantial Transverse Momentum}

\author{Cari Cesarotti}
\email{ccesarotti@g.harvard.edu}
\affiliation{Department of Physics, Harvard University, Cambridge, MA 02138}

\author{Yotam Soreq}
\email{yotam.soreq@cern.ch}
\affiliation{Theoretical Physics Department, CERN, Geneva, Switzerland}
\affiliation{Department of Physics, Technion, Haifa 32000, Israel}

\author{Matthew J. Strassler}
\email{strassler@physics.harvard.edu}
\affiliation{Department of Physics, Harvard University, Cambridge, MA 02138}

\author{Jesse Thaler}
\email{jthaler@mit.edu}
\affiliation{Center for Theoretical Physics, Massachusetts Institute of Technology, Cambridge, MA 02139}
\affiliation{Department of Physics, Harvard University, Cambridge, MA 02138}

\author{Wei Xue}
\email{wei.xue@cern.ch}
\affiliation{Theoretical Physics Department, CERN, Geneva, Switzerland}


\preprint{MIT-CTP/5044}
\preprint{CERN-TH-2018-188}

\begin{abstract}
We study dimuon events in 2.11\,fb\inv\ of 7 TeV $pp$ collisions, using CMS Open Data, and search for a narrow dimuon resonance with moderate mass (14--66 GeV) and substantial transverse momentum ($p_T$).
Applying dimuon $p_T$ cuts of 25 GeV and 60 GeV, we explore two overlapping samples: 
one with isolated muons, and one with prompt muons without an isolation requirement.
Using the latter sample requires information about detector effects and QCD backgrounds, which we obtain directly from the CMS Open Data.
We present model-independent  limits on the product of cross section, branching fraction, acceptance, and efficiencies.
These limits are stronger, relative to a corresponding inclusive search without a $p_T$ cut, by factors of as much as nine.
Our ``$p_T$-enhanced''  dimuon search strategy provides improved sensitivity to models in which a new particle is produced mainly in the decay of something heavier, as could occur, for example, in decays of the Higgs boson or of a TeV-scale top partner.
An implementation of this method with the current 13 TeV data should improve the sensitivity to such signals further by roughly an order of magnitude.
\end{abstract}

\maketitle

\tableofcontents

\section{Introduction}
\label{sec:intro}

The CERN Open Data portal \cite{CERNOpenData} aims to make data from the Large Hadron Collider~(LHC) publicly available as a long-term archive, with the first research-grade data from the CMS experiment released in 2014 \cite{CMSOpenData}.
In order to identify any issues that might interfere with their use by physicists of the future, it is important that open data frameworks be tested today.
There are good scientific motivations to make use of this resource \cite{CMS:OpenAccess}.
Open data makes it possible for scientists outside of the LHC collaborations to study Standard Model~(SM) processes that are not well modeled by Monte Carlo~(MC) generators, such as rare QCD backgrounds.
Together with detector-simulated samples, open data also makes it possible to test event analysis strategies that rely on a detailed understanding of detector effects.
The value of the CMS Open Data for exploratory studies of QCD has been demonstrated in \Refs{Tripathee:2017ybi,Larkoski:2017bvj}; see \Refs{Madrazo:2017qgh,Andrews:2018nwy,Andrews:2019faz} for machine-learning studies on detector-simulated CMS samples, \Refs{Kile:2017ryy,Kile:2017ccn,Kile:2017psu} for QCD studies on archival ALEPH data, and \Ref{CidVidal:2018blh} for a diphoton analysis with public LHCb data.

In this paper, we report the first utilization of the CMS Open Data in a search for Beyond the Standard Model~(BSM) phenomena.
We seek a new particle $V$ that decays promptly to dimuon pairs ($\mu^+\mu^-$) and is typically 
produced with substantial transverse momentum ($p_T$).
Our analysis is based on 2.11\,fb\inv\ of 7 TeV center-of-mass $pp$ collision events recorded by the CMS experiment during the first part of 2011 and made public through the CERN Open Data portal~\cite{CMS:DiMuonPrimary}.
We perform a narrow resonance search in the dimuon mass range $m_V \in [14,66]$ GeV and study the effect of modest cuts on $p_T$, namely $p^V_T>25 \GeV$ and 60 GeV; this approach (which we will refer to as ``$p_T$-enhanced'') could be applied to larger $p_T$ values as well, or alternatively to a cut on the $V$ boost factor $p^V_T/m_V$.
This type of search strategy was suggested some time ago~\cite{Strassler:MITBerkeley}, as one of several unconventional approaches for finding low-mass dilepton and diphoton resonances \cite{Strassler:KITP08}, but to our knowledge has never been carried out as a public analysis by the LHC collaborations. 
For this reason, the mass and $p_T$ regime we cover is relatively unexplored.
Moreover, our model-independent approach is complementary to highly targeted searches.

A low-mass, high-$p_T$ $V$ particle is well motivated.
LHC-accessible hidden sectors of new particles without SM gauge interactions can result in narrow neutral resonances appearing at any mass.
These scenarios are often called hidden valleys~\cite{Strassler:2006im} or dark sectors~\cite{Essig:2013lka,Alexander:2016aln}; famous examples arise in  twin Higgs models~\cite{Chacko:2005pe} and asymmetric dark matter~\cite{Nussinov:1985xr,Kaplan:2009ag}. 
Such hidden sectors would have small {\it direct} production rates at the LHC and at all previous colliders, but {\it indirect} production through the decay of a heavier particle may be much larger than direct production.
This heavier particle could be a known SM state ({\it e.g.}\ $W$, $Z$, Higgs boson, or top quark) or as yet undiscovered ({\it e.g.}\ a top partner that has escaped detection due to its exotic decays, or a heavy Higgs), and its production rate may be much larger at the LHC than at lower-energy colliders.
When indirect production via decay is common, a $V$ particle from a hidden sector may typically have moderate to high $p_T$, and a search involving a $p_T$ cut may preserve the signal while reducing SM backgrounds sharply.

For the specific case of a $V$ decaying to dimuons, Drell-Yan~(DY) and QCD backgrounds (including both real muons from hadron decays and fake muons) fall rapidly with the dimuon transverse momentum $\ptmm$.
In many models, the signal's $\ptmm$ spectrum is harder than that of the background, so even a rather modest cut on $\ptmm$ increases sensitivity.
While this is not the case for minimal dark photon models with kinetic mixing~\cite{Okun:1982xi,Galison:1983pa,Holdom:1985ag,Pospelov:2007mp,ArkaniHamed:2008qn,Bjorken:2009mm} (see related discussion in \Ref{Hoenig:2014dsa}), it is common to any scenario where $V$ is produced from the decay of a heavier state.
It is also the case, for example, in the SM search for $h \to \mu^+ \mu^-$, where $\ptmm$ is used to define event categories \cite{Aaboud:2017ojs} or as part of a multivariate discriminant \cite{Sirunyan:2018hbu}.

In addition to having substantial $p_T$, the $V$ may often be produced in association with other hidden sector particles~\cite{Strassler:2006im,Han:2007ae}, whose decay products might be clustered together in the detector.
This clustering could significantly reduce the efficiency of any lepton isolation cut on the signal.
To ensure sensitivity to the broadest range of models, one may wish to relax or drop the tight isolation criteria that are usually applied in dilepton searches, and instead reduce QCD backgrounds through a stringent impact parameter~(IP) cut to select real prompt muons.%
\footnote{Another potential failure mode for isolation can occur even with a single isolated highly boosted $V$, where the $V$ decay products ruin each other's isolation; cf.~early studies of lepton jets~\cite{Baumgart:2009tn,Cheung:2009su,Falkowski:2010cm,Falkowski:2010gv,Aad:2015sms} and photon jets~\cite{Dobrescu:2000jt,Larios:2001ma,Toro:2012sv, Draper:2012xt, Ellis:2012zp}.
One may evade this by excluding companion muons when imposing isolation.
We do not do so  here, but this should be implemented for any search targeting lower $V$ masses and/or higher $p_T$ cuts.}
With the availability of CMS Open Data and corresponding simulated samples, we can test the efficacy of an IP-cut-based search strategy and look for prompt but non-isolated dimuon resonances.

The use of a $p_T$ cut to separate a BSM signal from SM backgrounds has a long history.
In the LHC era, there has been intense interest in particles with a large boost, such that their decay products become highly collimated.
Searches for particles that decay to one or more highly boosted $W$/$Z$/Higgs bosons or top quarks have been widely proposed and carried out, for instance in \Ref{Chatrchyan:2012tw} for boosted $Z \to \mu^+ \mu^-$; see \Refs{Larkoski:2017jix,Asquith:2018igt} for recent reviews for boosted hadronic objects.
Searches for new particles that are produced with a high boost have also been proposed \cite{Strassler:2006im,Han:2007ae,Aguilar-Saavedra:2017zuc,Chakraborty:2017mbz,Aguilar-Saavedra:2017rzt,Aguilar-Saavedra:2018xpl,Collins:2018epr}, and although some have been implemented \cite{Chatrchyan:2012cg,Khachatryan:2015wka,Sirunyan:2018mgs}, there have been none to our knowledge in the purely dimuon  or dielectron channels.
Moreover, as we show here, enhanced sensitivity across the mass region of interest may be obtained even with moderate $p_T$ cuts, such that boost factors are typically much more modest.

In this paper, we present summaries of our \nameofsearch dimuon search results.
More details and additional results will be presented in future work.
In \Sec{sec:ZXsec}, we validate our use of the CMS 2011 dimuon data set by performing a measurement of the $Z$ boson cross section.
In \Sec{sec:Strategy}, we describe our dimuon resonance search strategy, with results shown in \Sec{sec:limits}.
Implications for various benchmark scenarios are sketched in \Sec{sec:app}, and we conclude in \Sec{sec:discussion}.

\section{Validation of the Dimuon Data Set}
\label{sec:ZXsec}

\subsection{Basic Selection Criteria}

Our analysis is based on the \texttt{DoubleMu} primary data set from CMS Run 2011A~\cite{CMS:DiMuonPrimary}, hereafter referred to as ``CMS11a'', 
and benefits from the excellent performance of the CMS muon system~\cite{Chatrchyan:2012xi,Chatrchyan:2013sba}.
We select events that pass the \texttt{HLT\_Mu13\_Mu8} ($\mu13\mu8$) high-level dimuon trigger, which nominally requires $p_T$ of $13~(8) \GeV$ for the leading~(subleading) muon.%
\footnote{The \texttt{DoubleMu} primary data set has 22 high-level trigger paths, none of which impose a muon isolation requirement, except \texttt{HLT\_DoubleMu5\_IsoMu5} which is not used here.}
To mitigate trigger threshold effects, we impose a further cut of $p_{T,1}^\mu > 15 \GeV$ on the leading muon and $p_{T,2}^\mu > 10 \GeV$ on the subleading muon, irrespective of their electric charge.
We also impose a pseudorapidity cut of $|\eta_\mu| < 2.1$, since the muon $p_T$ resolution degrades in the forward region.
We performed a validation study using the prescaled \texttt{HLT\_DoubleMu7}  ($\mu7\mu7$) trigger with a nominal threshold of $p_{T} > 7 \GeV$ on both muons.
After our baseline selection, the muon $p_T$ spectra from $\mu13\mu8$ and $\mu7\mu7$ are statistically equivalent, demonstrating that we are indeed working in the trigger plateau region.%
\footnote{Note that \Ref{Chatrchyan:2013tia}, which used the same trigger on the 2011 data set, applied looser requirements of $|\eta_\mu| < 2.4$ and $p_T > 14 \GeV$ ($9 \GeV$) on the leading~(subleading) muon.
In \Ref{Hoenig:2014dsa}, the results of \Ref{Chatrchyan:2013tia} were recast as a dark photon search, albeit with weaker limits than derived here due to the use of relatively coarse mass bins.}

For all of our analyses, we require that the muons pass the \emph{tight muon} selection criteria defined in \Ref{Chatrchyan:2012xi}.%
\footnote{This tight definition is taken from the 2010 CMS performance study \cite{Chatrchyan:2012xi}.  To our knowledge, there is no dedicated muon performance study from CMS on the 2011 data.  There is a study on the CMS 2012 data that recommends slightly different tight muon selection criteria \cite{CMS-DP-2014-020}, but that study is limited to muons with $p_T < 20 \GeV$.}
This means that the muon is reconstructed both as a ``global muon'' with the fit yielding $\chi^2/\text{d.o.f.} < 10$ and as a ``tracker muon'' with more than 10 inner-tracker hits.
As a baseline IP requirement, the reconstructed muon tracks must intersect the primary vertex within $d_0 < 2$\,mm in the $x$--$y$ plane and $z_0 < 10$\,mm in the $z$ direction.

We now present two validation studies of the CMS11a $\mu13\mu8$ trigger stream.
These same baseline requirements will be used in our dimuon search in \Sec{sec:Strategy}.

\subsection{Comparison to Monte Carlo Samples}

\begin{figure}[t]  
\begin{center}  
\includegraphics[width=0.99\columnwidth]{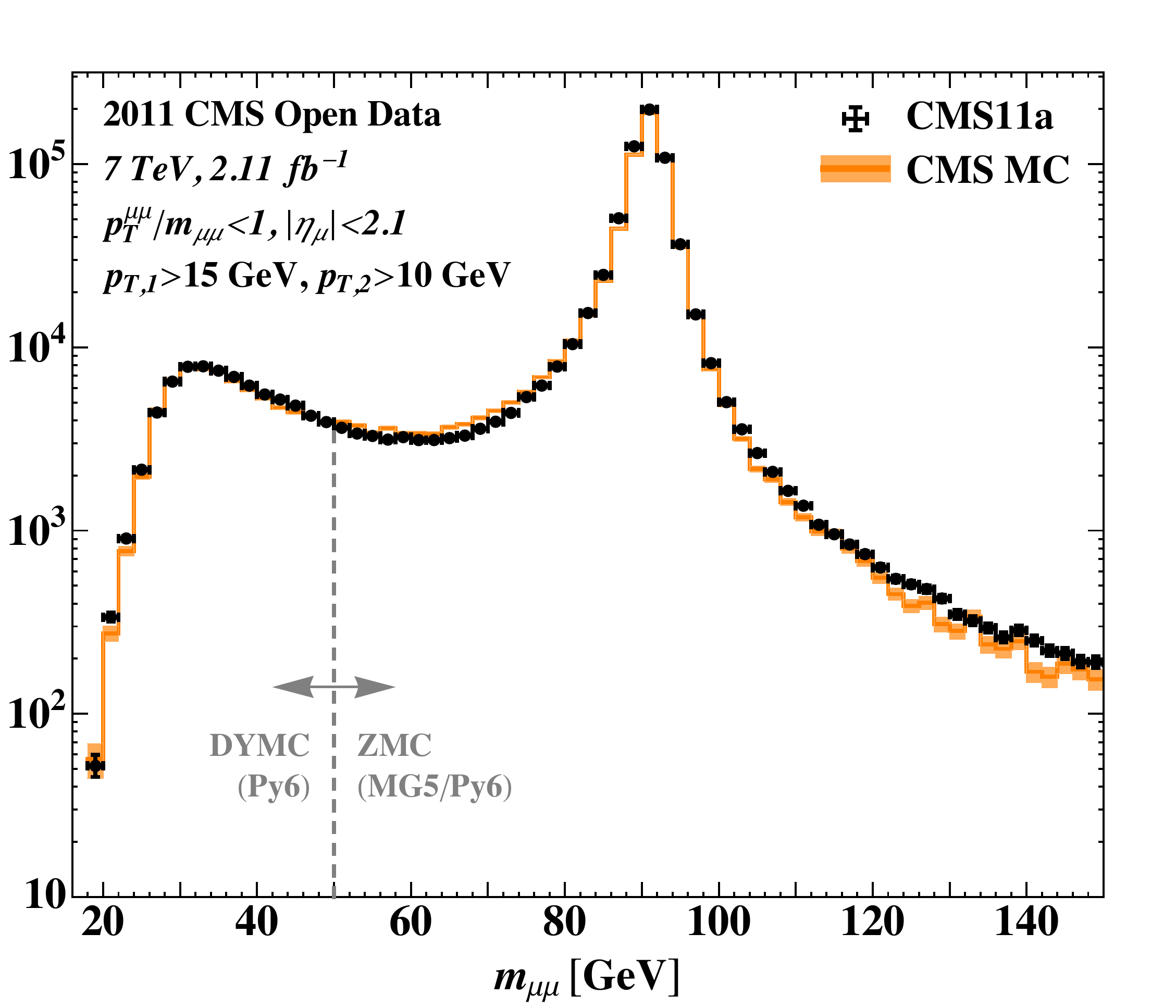} 
\includegraphics[width=0.99\columnwidth]{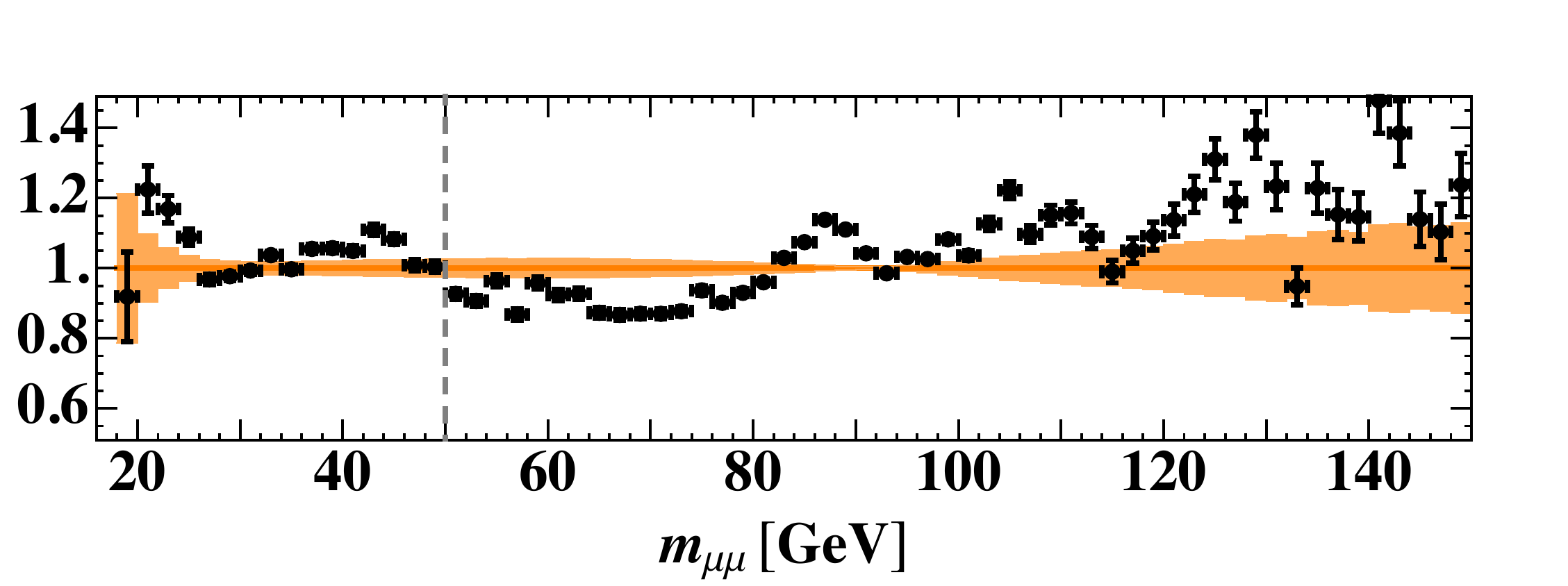} 
\end{center}
\caption{Dimuon mass spectrum for the CMS11a data set (black dots), compared to a combination of two MC samples (orange histogram) provided with the CMS Open Data:
DYMC (\textsc{Pythia} 6 with tune Z2) below 50 GeV and ZMC (\textsc{MadGraph} 5 interfaced with \textsc{Pythia} 6) above 50 GeV; see text for further details.
The normalization of the two MC samples is floated separately.
We require $\ptmm/m_{\mu\mu}<1$ because the DYMC sample does not have matrix element/parton shower matching; see the main text.
Statistical uncertainties on the data (MC samples) are shown as black error bars (orange shading). 
}
\label{fig:Fig1}
\end{figure}

The first validation study, shown in \Fig{fig:Fig1}, involves comparing the opposite-sign dimuon spectrum in the CMS11a data set with MC samples provided by CMS, which are generated using the CMS GEANT4-based \cite{Agostinelli:2002hh} detector simulation.
We impose isolation cuts on the muons to reduce QCD backgrounds to negligible levels; more details are given in \Eq{eq:Icomb_def} below.

In the mass range $m_{\mu\mu} > 50 \GeV$, we compare to a $Z$-pole Monte Carlo sample~(ZMC)~\cite{CMS:ZMC} obtained from \textsc{MadGraph} 5 v1.1.0 \cite{Alwall:2011uj} interfaced with \textsc{Pythia} 6.4.25~\cite{Sjostrand:2006za}%
\footnote{Since the information provided with \Ref{CMS:ZMC} (and other similar MC samples) does not specify the \textsc{Pythia} version used for event generation, we cite version 6.4.25, which has tune Z2 as an official option.  An earlier version might have been used, with the tune Z2 settings.}
with tune Z2~\cite{Field:2011iq} and TAUOLA 2.4~\cite{Jadach:1993hs}, adjusting the ZMC normalization to match the $Z$ boson peak in CMS11a. 
In the mass range $m_{\mu\mu} \in [10,50] \GeV$, we compare CMS11a to a DY Monte Carlo sample (DYMC)~\cite{CMS:DYMC} obtained from \textsc{Pythia} 6 with tune Z2, adjusting the DYMC normalization to match the top of the trigger turn-on curve around 30 GeV.  
We also impose an unusual {\it upper} bound of $\ptmm/m_{\mu\mu}<1$ on the data and MC events, because the DYMC sample, lacking parton shower/matrix element matching,  underestimates the high-$p_T$ tail of the data below $m_{\mu\mu}< 50 \GeV$.%
\footnote{We must rely on the unmatched \textsc{Pythia}-only DY sample here, because the 2011 CMS Open Data release provided \textsc{MadGraph}/\textsc{Pythia} matched samples for DY plus $\{1,2,3,4\}$ jets \cite{CMS:DY1JetMC,CMS:DY2JetMC,CMS:DY3JetMC,CMS:DY4JetMC} but not for DY plus 0 jets.  This highlights the importance of stress-testing archival data strategies, to ensure that relevant information is not inadvertently omitted.}

\begin{table*}[t]
\centering
\begin{tabular}{ r @{$\quad$} r @{$\quad$} r } 
 \hline
 \hline
  & Dimuon Events \\
\hline
CMS11a $\mu13\mu8$ & 6,241,576 & \\
Baseline Acceptance  &  \multirow{2}{*}{2,961,681} \\
($p_{T,1}^\mu > 15 \GeV$, $p_{T,2}^\mu> 10 \GeV$, $|\eta_\mu| < 2.1$)  \\
 Tight Muon Cuts  &  \multirow{2}{*}{2,155,900} \\
 ($\chi^2/\text{d.o.f.} < 10$, $n_{\rm hit} \ge 10$, $d_0<2$ mm, $z_0<10$ mm) \\
\hline
\hline
 & OS Events & SS Events \\
 \hline
 Opposite Sign vs. Same Sign & 1,895,756 & 260,144 \\
 $Z$-mass Region ($m_{\mu\mu}\in [60,120] \GeV$) & 794,623 & 30,105\\
 $p_{T}^\mu>20\GeV$ & 699,270 & 9,726\\
 Muon Isolation ($I_{\rm comb}<0.15$) & 642,219 & 78\\
 \hline
 \hline
 \end{tabular}
\caption{Cut flow for the analysis of $\sigma_{Z\mu\mu}$ using the CMS11a data set.}
\label{tab:ZXsecCut}
\end{table*}

The CMS11a data set and the DYMC/ZMC samples show fairly good agreement in \Fig{fig:Fig1}, including the shape of the $Z$ pole and the shape of the trigger turn-on region.
Below the $Z$ pole, disagreements are mostly within the expected theoretical uncertainties of the simulations, which are of order $\alpha_s\sim$ 10\%--20\%.
Where the DYMC and ZMC samples meet at $m_{\mu\mu} = 50 \GeV$, there is a small mismatch, again within the expected theoretical uncertainties of the simulations.
(Strictly speaking, the DYMC and ZMC samples are defined by the generator-level $m_{\mu\mu}$, not the reconstructed $m_{\mu\mu}$, so near $m_{\mu\mu} = 50 \GeV$ the DYMC and ZMC curves in \Fig{fig:Fig1} actually include a few events from the other sample.)
Above the $Z$ pole, the ZMC sample lacks the $t\bar t$ background present in data at high mass.
The $Z$-pole region shows that the ZMC underestimates the width of the $Z$ resonance in data, a known effect~\cite{Chatrchyan:2012xi}.
While CMS has documented three different methods (MuScleFit, Rochester, and SIDRA) to correct the MC resolution~\cite{CERNOpenDataMuonRecommendations}, details are not publicly available for the reprocessed CMS Open Data samples.
(Implementing the ``Summer11'' SIDRA correction~\cite{CMSSIDRA} on the ``Summer11LegDR'' ZMC sample leads to oversmearing of the $Z$ peak.)
We have no independent way to determine the corrections, but fortunately we will not need them elsewhere.
Within these limitations, the general agreement provides confidence that our data sample is in accord with expectations.

In carrying out this check, we should have first applied a scale factor correction on the muon $p_T$ to the data.
This scale factor is a function of $p_T$, $\eta$, and azimuthal angle $\phi$.  
However, this information for the CMS11a data set is not yet public and we are therefore unable to use it directly.
We can obtain some partial information as follows.
A study in \Ref{Chatrchyan:2012xi} shows how the uncorrected $Z$ mass, as a function of the charge-weighted muon azimuthal angle, varies by $\pm 0.6\%$ in 2010 data.  
Since we find that the corresponding variation in the CMS11a data set is much smaller, we infer that improved calibrations were applied to it.
We also find that the $J/\psi$ mass in our sample varies by less than $0.3\%$ for $|\eta|<2.1$.
In summary, we find evidence that the largest variations in the scale factor have already been corrected in CMS11a, and that any residual corrections to be accounted for are far below the 1\% level.

Meanwhile, the recommendation from CMS for the 2011 data is that, if unable to apply muon scale factor corrections, one should take the scale factor to be $1.000 \pm 0.002$  \cite{CERNOpenDataMuonRecommendations}.
This $0.2\%$ uncertainty has a negligible effect on the cross-checks in this section, so we do not account for it.

\subsection{Extracting the $Z$ Boson Cross Section}

For a second validation study, we extract the cross section $\sigma_{Z\mu\mu}$ for $Z$ bosons decaying to muons using the CMS11a data set.
Our analysis is modeled on the CMS measurement of $\sigma_{Z\mu\mu}$ on 36 pb\inv\ of 2010 data \cite{CMS:2011aa} (see also \Ref{Khachatryan:2010xn}).
We impose the same kinematic cuts:  $p^\mu_T> 20 \GeV$ and $|\eta_\mu| < 2.1$ on both muons, and $m_{\mu\mu} \in [60,120] \GeV$.
This ensures that our acceptance, and the theoretical SM cross section in the $Z$-mass window, should match \Ref{CMS:2011aa}, though the trigger and isolation criteria are different.
In \Tab{tab:ZXsecCut}, we show the number of dimuon events that pass these cuts, separated by whether the two leading muons have charges with the same sign (SS) or opposite sign (OS).

The quantity $\sigma_{Z\mu\mu}$ can be obtained from the number of $Z$ candidates, $N_Z$, via
\begin{align}
	\label{eq:ZXsec}
	\sigma_{Z\mu\mu} &= \sigma(pp\to Z+X) \, \BR(Z\to \mu^+\mu^-) \nonumber \\
	&= \frac{N_Z}{\cL \, A_Z \, \epsilon^Z_{\rm tr} \, \epsilon^Z_{\rm iso}  } \, ,
\end{align}
where $\cL$ is the integrated luminosity, $A_Z$ is the kinematic acceptance, $\epsilon^Z_{\rm tr}$ is the combined trigger/reconstruction efficiency for the $Z$ sample, and $\epsilon^Z_{\rm iso}$ represents the sample's isolation efficiency.
The central values and uncertainties for these quantities are summarized in \Tab{tab:AEff} and described briefly below.

Integrated luminosity information is provided with the CMS Open Data~\cite{CMS:2011Lumi}. 
To determine $\cL$, we sum over the luminosity blocks where the $\mu13\mu8$ trigger was active, obtaining 2.16\,fb$^{-1}$ delivered and 2.11\,fb$^{-1}$ recorded for CMS11a.%
\footnote{Strangely, there are 7 luminosity blocks where the recorded luminosity is zero, despite the fact that they contain a total of 17 events where the $\mu13\mu8$ trigger fired.  Removing these events has a negligible impact on our results.}
CMS quotes a 2.2\% luminosity uncertainty for 2011~\cite{CMS-PAS-SMP-12-008}, and we take this as a systematic uncertainty.

Though we cannot cross-check the luminosity uncertainty independently, we did verify that when we break the data into subsets, the number of events in the $Z$ boson peak divided by the integrated luminosity is nearly constant.  
The same is true for the number of non-$Z$ Drell-Yan events, which is a further check that the $\mu13\mu8$  trigger functioned stably during the run.
That said, there is some jitter in these ratios, of order 2\%.
We have no information about the source of this jitter, which could stem from the luminosity measurement, the trigger/reconstruction efficiency, or other sources.
To be conservative we assign this uncertainty to the trigger/reconstruction efficiency; see below.

\begin{table}[t!]
\centering
\begin{tabular}{ r @{$\quad$}  c @{$\quad$} c} 
 \hline \hline
 & Central Value & Uncertainty \\
 \hline
 $\cL$ & 2.11\,fb\inv & 2.2\% \\
 $A_Z$ & 0.392 & 2.4\% \\
 $\sqrt{\epsilon^Z_{\rm tr}}$\ ({\it i.e.}\ per muon) & 0.924 & 2.4\%\\
 $\sqrt{\epsilon^Z_{\rm iso}}$\ ({\it i.e.}\ per muon) & 0.966 & 1.5\%\\
 Background & --- & 1.0\% \\
 \hline
Combined ($\cL \, A_Z \, \epsilon^Z_{\rm tr} \, \epsilon^Z_{\rm iso}$) & 0.659\,fb\inv & 5.3\% \\
 \hline
 \hline
 \end{tabular}
\caption{
  A summary of the acceptance and efficiency factors for the $\sigma_{Z\mu\mu}$ analysis. We show single-muon trigger/reconstruction and isolation efficiencies since these are what we actually compute, using a combination of MC-based and  data-driven methods.}
\label{tab:AEff}
\end{table}

Using the ZMC sample, we find a kinematic acceptance factor of $A_Z = 39.2\%$, to be compared with the $39.8\%$ in~\Ref{CMS:2011aa}; we take the relative $1.5\%$ discrepancy as a systematic uncertainty.
There is also a 1.9\% theoretical uncertainty on $A_Z$ noted in \Ref{CMS:2011aa}, which we combine in quadrature for a total $A_Z$ uncertainty of 2.4\%.

Since~\Ref{CMS:2011aa} uses a single-muon trigger (whereas we use a dimuon trigger), and applies cuts for muon quality and isolation that differ from ours, we must determine the corresponding efficiencies ourselves; details on this procedure will be presented in future work.
For the trigger/reconstruction efficiency $\epsilon^Z_{\rm tr}$, we must rely on truth information from the ZMC sample, but the result we find can be cross-checked against 2011 CMS estimates, such as found in \Ref{CMSMuonTwiki}; these show that, for the single-muon efficiency, MC and data agree to within 2\%, which we take to be a systematic error.
We combine this in quadrature with the uncertainty inferred from the jitter in the ratio of $Z$ boson events to recorded luminosity, 2\% on the {\it dimuon} efficiency (1.4\% per muon), giving a total uncertainty on the single-muon efficiency of 2.4\%.

To impose isolation, we require $I_{\rm comb}<0.15$ for each muon, where the combined isolation variable is 
\begin{equation}
\label{eq:Icomb_def}
I_{\rm comb} = \frac{ \big(p_T^{\rm track} + E_T^{\rm ECAL} + E_T^{\rm HCAL} \big)_{R < 0.3} } {p_T^\mu},
\end{equation}
where the numerator is the sum of the transverse momenta of all tracks within a cone of radius $R=0.3$ around the muon, together with the transverse energy of all ECAL (electromagnetic calorimeter) and HCAL (hadronic calorimeter) deposits within the same cone, without removing double counting (see \Ref{Chatrchyan:2012xi}).
To determine $\epsilon^Z_{\rm iso}$, we use multiple methods, including truth information from ZMC and a tag-and-probe analysis on the CMS11a data, and these agree to within 1\%.
To be conservative we take a 1.5\% systematic uncertainty.

This analysis is essentially background free.
This can be seen, for instance, in the CMS DY study \cite{Chatrchyan:2013tia}, where backgrounds from $Z\to\tau\tau$, $t\bar t$, $WZ$, $ZZ$, and QCD ({\it i.e.}\ real and fake muons from all hadronic sources) together add up to less than 1\% of the signal.
This can be checked by a direct calculation, except for the QCD background, which we probe using SS muon events; from \Tab{tab:ZXsecCut} we see that they are removed efficiently by the isolation cut.
Combining the uncertainties from \Tab{tab:AEff} in quadrature leads to a relative uncertainty of approximately 5\%.

Inserting $N_Z$ from \Tab{tab:ZXsecCut} into \Eq{eq:ZXsec}, we find 
\begin{align}
\sigma_{Z\mu\mu}  = ( 974\pm1\pm52)\,{\rm pb}
\end{align}
in the $Z$-mass window of 60--120 GeV, where the first uncertainty is statistical and the second is from the uncertainties in \Tab{tab:AEff}.
This agrees with
the next-to-next-to-leading-order SM prediction of $970\pm 30\,$pb quoted in \Ref{CMS:2011aa} (obtained from FEWZ~\cite{Gavin:2010az} and MSTW08~\cite{Martin:2009iq}), 
the measured value of $968\pm44\,$pb in~\Ref{CMS:2011aa} ($974\pm44\,$pb with electron/muon averaging),
and the 2011 CMS result of $986\pm31$\,pb~\cite{Chatrchyan:2013tia}.

\section{Resonance Search Strategy}
\label{sec:Strategy}

We now describe our analysis strategy for setting new bounds on $V$ production.
Our results are largely model independent, up to subtleties described below.
The overall methodology is straightforward.
Taking events in the $\mu13\mu8$ trigger stream, we impose minimal additional cuts on the $\eta$ and $p_T$ of the muons.
We then define separate {\it isolated} and {\it prompt} samples that overlap but are useful for different classes of signal models.
We finally impose three different cuts on the dimuon transverse momentum $\ptmm$ to isolate boosted kinematics.
Within these samples (six in total), we search for a narrow bump, with a width appropriate to the CMS dimuon mass resolution and a Crystal-Ball-like line shape.
We employ a profiled-likelihood method using approximate formulas from~\Ref{Cowan:2010js}, with certain details motivated by~\Ref{Williams:2017gwf}.

\begin{table*}[t!]
\centering
\begin{tabular}{r @{$\quad$} r @{$\quad$} r} 
 \hline
 \hline
  & Dimuon Events & \\
 \hline
 Baseline Acceptance and Tight Muons Cuts &  \multirow{3}{*}{2,155,900}  & \\
 ($p_{T,1}^\mu > 15 \GeV$, $p_{T,2}^\mu> 10 \GeV$, $|\eta_\mu|<2.1$, \\
  $d_0<2$ mm, $z_0<10$ mm, to match \Tab{tab:ZXsecCut}) & \\
 \hline
 Search Region & \multirow{3}{*}{561,364} & \\
 (OS, $m_{\mu\mu} \in [11,83]  \GeV$, & \\
 $d_0<250\,\mu$m, $z_0<2000\,\mu$m)  & \\
\hline
\hline
&  Isolated Sample & Prompt Sample \\
&  ($I_{\rm comb} <0.15$) & ($d_0<100\,\mu$m)  \\

 \hline
 \hline
 $\ptmm>0$  & 188,924 & 412,002 \\
 $\ptmm>25\,$GeV  &46,798 & 91,264 \\
 $\ptmm>60\,$GeV  &7,668 & 11,208 \\
\hline
 \hline
 \end{tabular}
\caption{Cut flow for the $V\to\mu^+\mu^-$ search, illustrating the number of CMS11a events surviving various requirements.  The population of the six signal regions is shown.
}
\label{tab:BSMFlow}
\end{table*}

\subsection{Defining Isolated and Prompt Samples}

The initial event selection mirrors that of \Sec{sec:ZXsec}, \Tab{tab:ZXsecCut}.
As summarized in \Tab{tab:BSMFlow}, we place $p_T$ cuts of 15~(10) GeV on the leading~(subleading) muon to ensure that we are above the $\mu13\mu8$ trigger threshold.
We require these two muons to satisfy $|\eta_\mu|<2.1$ because of the degraded $p_T$ resolution at forward angles, and we demand that they satisfy the transverse and longitudinal IP requirements of $d_0<2$\,mm and $z_0<10$\,mm.
Next, we tighten the IP cuts to $d_0<250 \, \mu$m and $z_0< 2000 \, \mu$m, and limit ourselves to OS events in the mass window  $m_{\mu\mu} \in [11,83] \GeV$, allowing for searches in the mass range $m_{\mu\mu} \in [14,66] \GeV$.

We then define two overlapping samples for study:
\begin{enumerate}
\item an {\it isolated sample}, where the two leading muons satisfy an isolation requirement of $I_{\rm comb} <0.15$
[defined in \Eq{eq:Icomb_def}], which dramatically suppresses the QCD background; and 
\item a {\it prompt sample}, where no isolation cut is imposed but the transverse IP cut on the two leading muons is tightened further to $d_0<100 \, \mu$m, substantially reducing the QCD background and leaving it comparable to the irreducible DY background. 
\end{enumerate}
From the ZMC and DYMC samples, and cross-checking using data,
we infer that this tighter IP cut in the prompt sample accepts $\geq97\%$ of typical prompt signals, an effect we correct for later.%
\footnote{The $Z$ sample of \Sec{sec:ZXsec}, and any high-$p_T$ sample of DY with isolation imposed, are almost free of QCD contamination.
This can be inferred from the number of SS dimuon events and from lack of a tail in the IP distribution.
In these nearly pure samples of prompt dimuons, which closely resemble our signals, we can directly estimate the relevant efficiency by counting events as a function of the IP cut.}
 Note that access to the CMS Open Data was essential for validating the prompt sample, since it involves QCD backgrounds whose magnitude cannot be precisely predicted {\it a priori}, as well as detector effects related to the IP resolution.
(Though not directly comparable, one can also infer the potency of the IP cut to reduce QCD backgrounds from Fig.~7b of \Ref{Chatrchyan:2012xi}.)

As control samples, we take SS muons separated into prompt and non-prompt subsamples, and OS muons where we reverse either the isolation cut or the tighter IP cut.
Nothing striking appears in these samples, which adds confidence that any features observed in the signal samples are not a result of kinematic sculpting.

\begin{figure*}[t]  
\begin{center}  
\leavevmode
\vskip 0.0in
\includegraphics[width=\columnwidth]{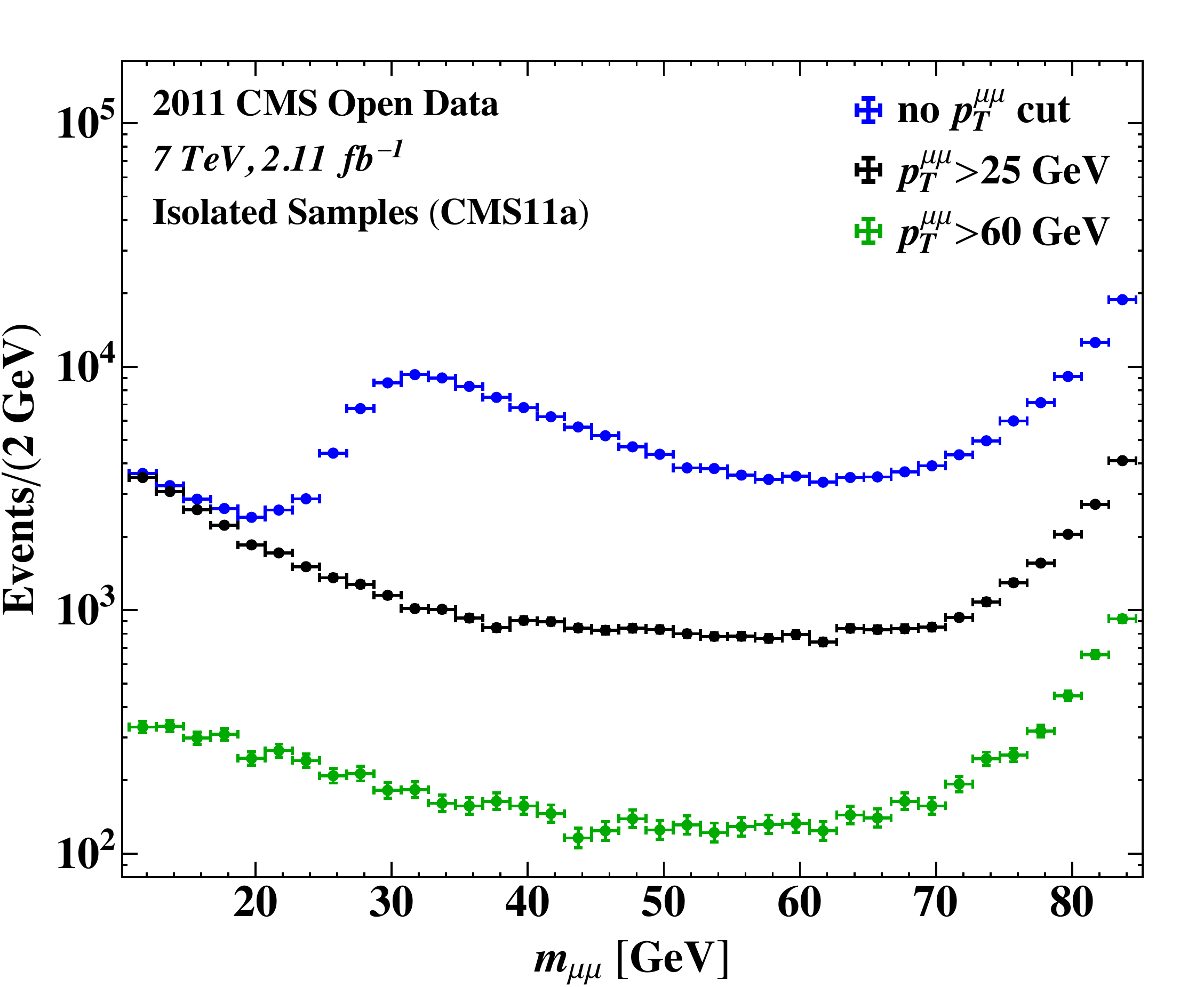} \
\includegraphics[width=\columnwidth]{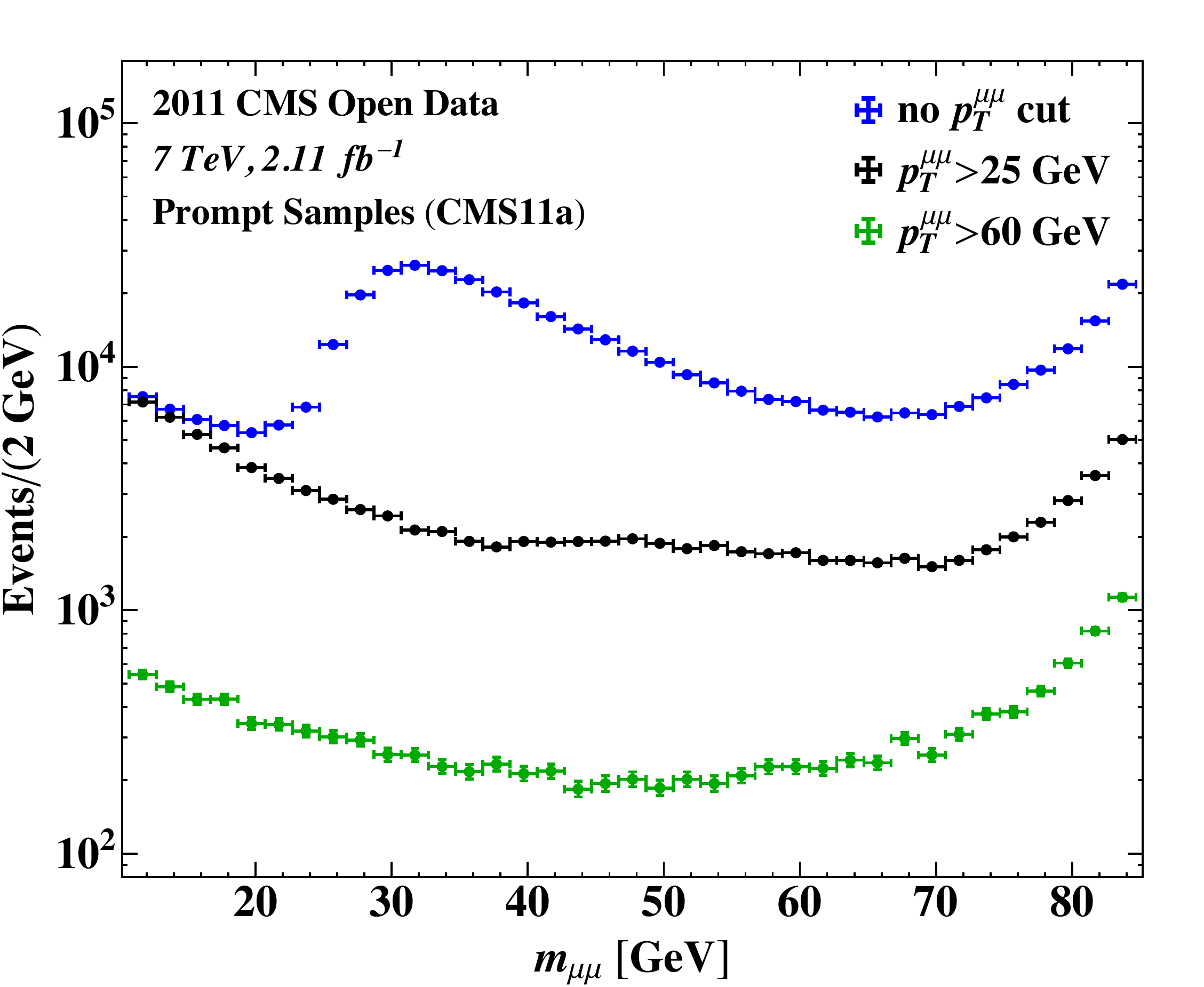} 
\end{center}
\vskip -0.20in  
\caption{The dimuon mass spectrum in 2 GeV bins for the (left) isolated and (right) prompt samples.
The distributions are shown for no $\ptmm$ cut (upper curve, blue) and for $\ptmm$ cuts of 25 GeV (middle, black) and 60 GeV (lower, green).
The trigger threshold, which dominates the inclusive sample, becomes irrelevant as $\ptmm$ cuts are applied.}  
\label{fig:DataByPtCut}
\end{figure*}

Finally, within the isolated and prompt samples, we consider additional subsamples, defined inclusively, in which we impose a $\ptmm$ cut.
The sequence of cuts is chosen  based on a principle: {\it a signal at the expected exclusion level (2$\sigma$) of one $p_T$ cut should be discoverable (5$\sigma$) following the next, tighter $p_T$ cut, assuming both cuts have identical signal acceptance.}
As we will see explicitly in \Sec{sec:app}, the latter assumption is more sensible than it might at first appear; it is often the case that a hard $p_T$ cut has high 
(60\%--100\%) signal acceptance relative to the next-hardest cut.
Based on this principle, we take three $\ptmm$ cuts of $\{0 \GeV, 25 \GeV, 60 \GeV\}$, which reduce the background at each step by approximately a factor of $(5/2)^2$, as shown in \Fig{fig:DataByPtCut}.
(One should continue this procedure as far as possible, but the next natural cut at $\ptmm>100$ GeV leaves little data in CMS11a; we do not study it here.)  
Of course, this reduction factor is not entirely uniform across the dimuon spectrum; because of the trigger's impact, the factor for the $\ptmm>25 \GeV$ cut is $\approx1$ below $25$ GeV, rising to $\mathcal{O}(10)$ just above this range.

The behavior seen in \Fig{fig:DataByPtCut} also explains why we chose, in this study, to make a cut on $\ptmm$ rather than on the dimuon boost $\ptmm/m_{\mu\mu}$.
The backgrounds at fixed $p_T$ are, somewhat accidentally, rather flat across this mass range; they are relatively easy to fit and we obtain bounds that are fairly uniform as a function of mass.
By contrast, backgrounds at fixed boost drop much more sharply across this mass range, complicating the fitting procedure.

\subsection{Resonance Line Shape}
\label{subsec:lineshape}

We next search for a bump, scanning across a range of values for the dimuon invariant mass $m_{\mu\mu}$.
As a potential signal, we assume a narrow resonance, with intrinsic width far smaller than the detector resolution.

The choice of line shape and resolution for our search requires some care, because the line shape for a signal is not model independent.
First, there is a radiative tail from QED emission off of the muons, whose precise form depends logarithmically on $m_V$.
Next and more importantly, the muon $p_T$ resolution, and therefore the dimuon mass resolution, is a function of the $p_T$ and especially the $\eta$ of the muons.
Since different models produce different muon $p_T$ and $\eta$ distributions, their line shapes will have different widths.
Finally, even at fixed $p_T$ and $\eta$, both the CMS11a data and the corresponding MC samples indicate that the resolution has small non-Gaussian tails.

In order to understand the CMS dimuon mass resolution $\rho_V(m_{\mu\mu})\approx \zeta\, m_{\mu\mu}$, which depends on $\eta$ and $p_T$,
we have studied the kinematic dependence of the CMS muon momentum resolution, using the line shape of the $J/\psi$ in the CMS11a data.  
(Samples of other hadronic resonances are either less abundant or less pure.)
The excellent tracking granularity means the resolution on the dimuon opening angle is subdominant to the $p_T$ resolution of the individual muons.
Because of the trigger, these $J/\psi$'s are highly boosted, and so the muons are very close in $\eta$ and in roughly the same $p_T$ range.
Fitting the $J/\psi$ line shape with a Crystal Ball function to account for both the radiative tail and the resolution allows us to estimate the $p_T$ resolution as a function of $\eta$ for $p_T^\mu\in[10,25]$ GeV and, with larger uncertainties, for $p^\mu_T > 25$ GeV. 
We also have an estimate of the resolution from the CMS MC samples, where we can directly relate generator-level and detector-level $p_T$ values.
Comparing data and MC indicates that the MC underestimates the resolution in the real data by about 10\%, but the $\eta$ dependence is otherwise well modeled within the region $|\eta_\mu|<2.1$ of interest to us.
Therefore, in $(p_T,\eta)$ bins where the CMS11a $J/\psi$ sample is large, we use the results from our fit to the $J/\psi$ as our central value, and at higher $p_T$, where the $J/\psi$ sample is too small, we use the resolution found in the CMS MC samples, multiplied by 1.1, as our central value.
Convolving these results against typical signal distributions, we find that the resolution is in the range $\zeta\,\sim 1.1\%$ for low $\ptmm$, slowly increasing for higher $p_T$ signals to around $1.3\%$ for $\ptmm\sim 60$ GeV.

The uncertainty in the resolution is difficult for us to determine, since the current release of CMS Open Data does not provide any detailed information concerning muon resolution and its uncertainty.
It is recommended~\cite{CERNOpenDataMuonRecommendations}, when unable to apply resolution corrections in detail, to take a systematic uncertainty of $\pm 0.6 \%$ in the $p_T$ resolution.
This appears consistent with the uncertainties found in the most up-to-date public information from 2010 \cite{Chatrchyan:2012xi}.
The corresponding uncertainty of $\pm0.4\%$ on the dimuon mass resolution appears to be too large, based on our studies of the $J/\psi$ in the CMS11a sample, but since we cannot quantify this reliably, we follow the above recommendation.

As noted earlier, we also follow the CMS recommendation for our data set to take the scale factor on the muon $p_T$ to be $1.000\pm0.002$~\cite{CERNOpenDataMuonRecommendations}.
The resulting scale uncertainty on $m_{\mu\mu}$ of $0.14\%$ is approximately half the size of our bins, and $\sim{1}/{8}$ the size of our signal resolution.
We cannot model the scale factor uncertainty properly, since we have no information about the dependence of the scale factor on kinematic quantities, and thus no information about event-to-event correlations.
But a constant (event-independent) scale factor of 1.0014 would shift the dimuon mass by less than 100 MeV for $m_{\mu\mu}<66$ GeV, too small to affect our analysis, and an uncorrelated one would combine in quadrature with the uncertainty of $0.4\%$ in the resolution, leaving it unchanged to the available precision.
We consequently do not account for this uncertainty in our results.

\begin{table*}[t]
\centering
\begin{tabular}{ r @{$\quad$}  c @{$\quad$} c @{$\quad$} c} 
 \hline \hline
 & Central Value & Uncertainty & Incremental Effect on Expected Limit\\
 \hline
 Resolution & 1.1\% (1.3\%)  & 0.4\% &10\% (7\%)  (profiled) \\
\hline
 Line Shape Modeling & 1 & 5\% &  \\
 $\cL$ & 2.11\,fb\inv & 2.2\% & {0.6\%} (multiplicative) \\
$\epsilon^V_{\rm IP}$ (prompt sample only) & 0.97 & 1.5\% & \\
 \hline
 \hline
 \end{tabular}
\caption{A summary of the systematic uncertainties on our fitting results, showing the size of the uncertainty and the effect on our limits.
We profile explicitly over the resolution.
The latter three uncertainties, which are essentially uniform across our mass range, are combined together in quadrature and assessed, after the resolution profiling, using the multiplicative approach in \Eq{eq:rescaling_trick}.
(Because the uncertainty in $\epsilon^V_{\rm IP}$ is so subdominant, its presence in the prompt samples does not alter the incremental effect on the expected limits.)
}
\label{tab:SysUnc}
\end{table*}

The appropriate line shape has a Gaussian core and a radiative tail.\footnote{A Gaussian line shape, without accounting for the radiative tail, gives limits  $\sim$ 10\%--15\% smaller than those presented below.}
The most important role of the tail is to deplete signal from the Gaussian core, so it is important that its integral be approximately correct in order that the core be properly normalized.  
We cannot determine this entirely from data, because the radiative tail from the $J/\psi$, the cleanest resonance, disappears under the continuum background. 
For this reason, a MC-based approach for modeling this well-understood QED phenomenon is more accurate.

We therefore first generate a high-statistics sample of $V$ decays with \textsc{Pythia}~8.235~\cite{Sjostrand:2014zea}, in a specific model for the $V$ kinematics (model M1 defined in \Sec{subsec:viadecay}),  for $m_V=3.1$ GeV and for $m_V\in[14,66]$ GeV.
This generator includes photon final state radiation (FSR), so the dimuon mass distribution has a tail below the delta function spike at $m_V$, whose size depends on $\log(m_V/m_{\mu})$.
We then smear this result with a Gaussian, applied event by event according to the $p_T$- and $\eta$-dependent single-muon $p_T$ resolution obtained above. 
The amount of smearing is chosen so that for $m_V=3.1$ GeV we reproduce the desired $p_T$- and $\eta$-dependent resolution in the core of the $J/\psi$ peak to within 0.05\%, much smaller than the uncertainties on the resolution of $0.4\%$.
We then apply the same procedure for other $m_V$ to obtain a predicted line shape (with a slow dependence on $m_V$ and specific to model M1) for the central value of the resolution.
We repeat the procedure, increasing or decreasing the smearing by an amount that is independent of $p_T$ and $\eta$, to obtain other choices of resolution that we need later in our statistical analysis.%
\footnote{The CMS11a data and MC reveal subtleties in the efficiency for muon reconstruction when a hard muon overlaps with a hard FSR photon. But this issue only affects dimuons far into the radiative tail, and does not impact our results.} 

Our generated statistics are high enough that we may use the smeared MC as our prediction.
As a check, we studied smoothing our prediction by fitting it with a single- or double-shouldered Crystal Ball function.
These fits give results that differ by up to 3\% on expected limits and up to 6\% on observed limits, but this is caused by an imperfect fit in the peak region, not by low MC statistics on the tails.
Nevertheless our prediction has intrinsic uncertainties, both from the modeling of photon FSR and from the fact that detector effects produce slightly non-Gaussian smearing, but these are common to all samples and vary little if at all with $m_V$.
We  associate to these effects a  5\% conservative Gaussian uncertainty in the best fit signal strength that affects all samples and masses uniformly.
The impact on our 95\% confidence upper limits is then very small, as we will see below.

\subsection{Systematic Uncertainties}
\label{subsec:systematics}

Our results include systematic uncertainties associated with the four effects in \Tab{tab:SysUnc}.
For the dimuon resolution, we take central values of $\zeta = 1.1\%$ for the $\ptmm > \{0 \GeV,25 \GeV\}$ samples and $\zeta = 1.3\%$ for $\ptmm > 60\GeV$, and we profile over the $\pm0.4\%$ resolution uncertainty as described in \Sec{subsec:limit_procedure} below.
As discussed further in \Sec{subsec:use_results}, we externalize the uncertainties associated with the acceptance and trigger/reconstruction efficiencies, since they are model dependent.

The three remaining uncertainties are from line-shape modeling, luminosity, and (for the prompt sample only) IP cut efficiency.
The latter two effects have an obvious multiplicative impact on the limit.
Less obvious is that the line-shape uncertainty also has a dominantly multiplicative effect.
The reason is that, as far as fitting the signal is concerned, changing the tail of the line shape primarily changes the normalization of the Gaussian-like core.
While it is possible to profile over these multiplicative uncertainties, we can use a simpler rescaling procedure since these multiplicative effects are relatively small.

Let the signal strength $\mu = \xi \, \nu$ be multiplicatively proportional to a dimensionless quantity $\xi$ with Gaussian uncertainty $\delta \xi$ and central value $\xi_0$.
Assume further that the log-likelihood profiled over all other quantities is effectively Gaussian, such that the quantity $\nu$ can be treated as having Gaussian uncertainty $\delta \nu$ and central value $\nu_0$:
\begin{equation}
\label{assumedLL}
\Lambda(\xi,\nu) \approx \Lambda_{\min} + \left(\frac{\xi-\xi_0}{\delta \xi} \right)^2 + \left(\frac{\nu-\nu_0}{\delta \nu} \right)^2 \ .
\end{equation}
Marginalizing over $\xi$ and $\nu$, keeping $\mu$ fixed, and taking the $\delta \xi \ll  \xi_0$ limit,\footnote{Strictly speaking, we have to assume that $\delta \xi / \xi_0 \ll \sigma_0 / \mu$, which is a reasonable approximation when evaluating the 95\% $CL_s$ lower/upper limit.} \Eq{assumedLL} becomes
\begin{equation}
\Lambda(\mu) \approx \Lambda_{\min} + \left(\frac{\mu-\mu_0}{\delta \mu}\right)^2, \quad \delta \mu \approx \sigma_0  \left(1 + \frac{1}{2} \frac{\delta \xi^2}{\xi_0^2} \frac{\mu^2}{\sigma_0^2} \right),
\end{equation}
where $\mu_0 = \xi_0 \, \nu_0$ and $\sigma_0 = \xi_0 \, \delta \nu$.
Thus, the profiled log-likelihood is shallower than when $\delta \xi = 0$, increasing the size of the $\mu$ confidence intervals.
For instance, the expected 95\% $CL_s$ upper limit  increases by
\begin{equation}
\label{eq:rescaling_trick}
\mu_{95} \approx \mu_{95}^{\delta \xi = 0} \left( 1 + \frac{1}{2}  \frac{\delta \xi^2}{\xi_0^2}{\Delta \Lambda} \right),
\end{equation}
with $\Delta \Lambda = 3.84$.

Because the corrections from these multiplicative uncertainties are quadratic in $\delta \xi$, their effect on our results is small.
When combined in quadrature in \Tab{tab:SysUnc}, the line-shape uncertainty dominates, leading to a shift in the expected limits of around 0.6\%.
Note that \Eq{eq:rescaling_trick} is obtained \emph{after} profiling over the resolution uncertainty and background fit, which explains why the impact of the multiplicative corrections is diluted in this analysis.

\subsection{Procedure for Setting Limits}
\label{subsec:limit_procedure}

We use the following procedure to obtain limits on $V \to \mu^+ \mu^-$ production, with more justification presented below.
For each mass value, we select a window centered around $m_{\mu\mu}$ of width  $35\, \rho_V$, binning the data in  140 bins.
We then fit the mass spectrum within the window to a background model, with or without a signal (whose line shape is described in \Sec{subsec:lineshape}) added at the center of the window.
The background is modeled as a fifth-order polynomial, including all orders from $x^0$ to $x^5$, with six free parameters that we profile over.
The signal shape is as described in \Sec{subsec:lineshape}, with a resolution profiled over the above-mentioned 0.4\% uncertainty, treated as Gaussian.
(When profiling the resolution, we still keep the window size fixed to 35 times the central value of the resolution.)
Using the above signal shape and background model, we determine a $p$-value for rejecting the background-only hypothesis, and evaluate observed and expected 95\% $CL_s$ upper limits on the number of signal events.
The expected limit is determined from the Asimov data set~\cite{Cowan:2010js} in the standard way.
We incorporate various uniform systematic uncertainties, shown in \Tab{tab:SysUnc}, by adding them into the likelihood and computing their effects on the limits analytically (see \Eq{eq:rescaling_trick} above).

The choice of the above background fitting method is motivated as follows.
The available MC samples from CMS do not allow us to reliably predict the background in all relevant kinematic regions, so we cannot determine a fit function {\it a priori} over the whole mass range.
We therefore fit to the background locally in a window around each dimuon mass value, and we use a polynomial fit because of the somewhat intricate shape of the background.
The use of a polynomial background fit in a centered mass window was advocated for in \Ref{Williams:2017gwf} and employed in \Ref{Aaij:2017rft}.
In this approach, both the degree of the polynomial and size of the window (relative to the resolution $\rho_V$) must be chosen.

In order for the background to be well modeled by a polynomial, we should choose a high-order polynomial and a small mass window.
In particular, a window larger than roughly $m_{\mu\mu}/2$ covers so much of the data that it defeats the purpose of local fitting.
Because we center the mass window, adding an odd-order to an even-order polynomial has almost no effect on our results, as a parity-odd term is orthogonal to a Gaussian signal and nearly orthogonal to a more realistic signal with a radiative tail~\cite{Williams:2017gwf}.
We therefore consider odd-order polynomials of third order or higher (since a linear fit function gives bad fits with any reasonable choice of window), and windows no larger than $50 \, \rho_V$ (to be compared to $25 \,\rho_V$ recommended in \Ref{Williams:2017gwf}).

On the other hand, a mass window that is too small, or a polynomial that has too high an order, leads to a spurious ``ringing'' effect: a large excess at one mass can affect the fits at nearby masses, generating subsidiary correlated $p$-value spikes on either side of a real spike.
These correlated spikes, visible by eye, are also detectable through the distribution of spikes as a function of local $p$-value, and equivalently by unreasonably large global $p$-values relative to the maximum local $p$-value.
We find that avoiding the ringing effect requires a window of at least 25 (30)  $\rho_V$ for a cubic (quintic) polynomial.
Our results are stable for a range of window sizes above these values, except in the trigger turn-on region for the inclusive $\ptmm>0$ subsample, which we mask in the limits below.
A seventh-order polynomial appears to require a window too large for good fits.

Limits obtained using the quintic, with more nuisance parameters, are generally higher than those for the cubic.
We therefore use the quintic as the more conservative option, effectively soaking up the systematic uncertainty associated with the choice of background model by profiling over two additional parameters.
We retain the cubic as a cross-check, and we also check the stability of the limits using windows of $30\, \rho_V$ and $40\, \rho_V$.
In the spirit of \Ref{Williams:2017gwf}, we tested the impact of discretely profiling over the cubic and quintic models, finding results that were generally intermediate between those of the two polynomials taken separately.
Details and further justifications of our methods will be provided in future work, in which we also ``search'' for and observe, in the prompt sample, the SM meson decay $\eta\to\mu^+\mu^-$.

\section{Limits On Dimuons Using Cuts on Transverse Momentum}
\label{sec:limits}

\subsection{Search Results}
\label{subsec:searchresults}

%
\begin{figure*}[t]  
\begin{center}  
\leavevmode
\vskip 0.0in
\includegraphics[width=\columnwidth]{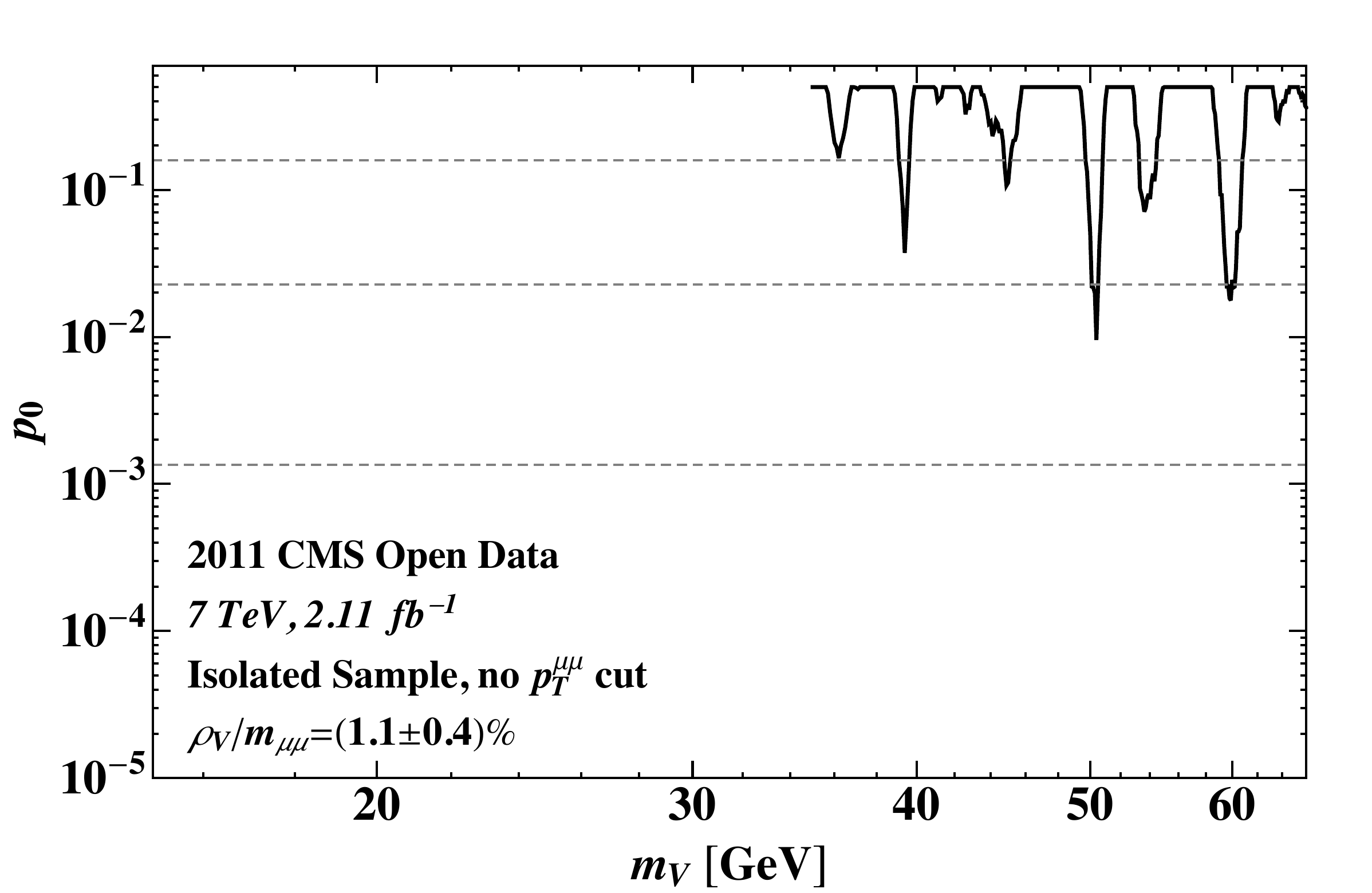} \
\includegraphics[width=\columnwidth]{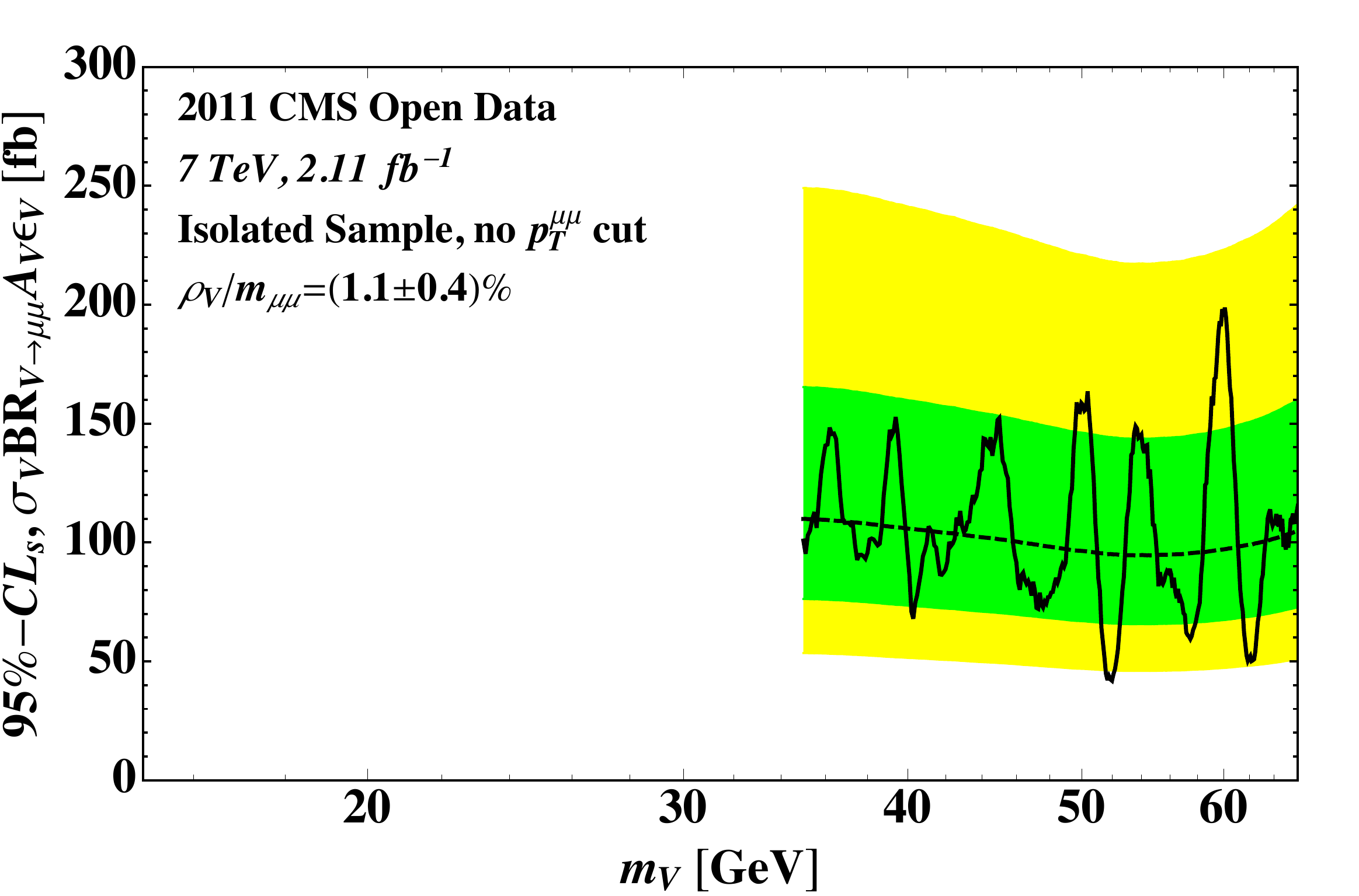} \\
\includegraphics[width=\columnwidth]{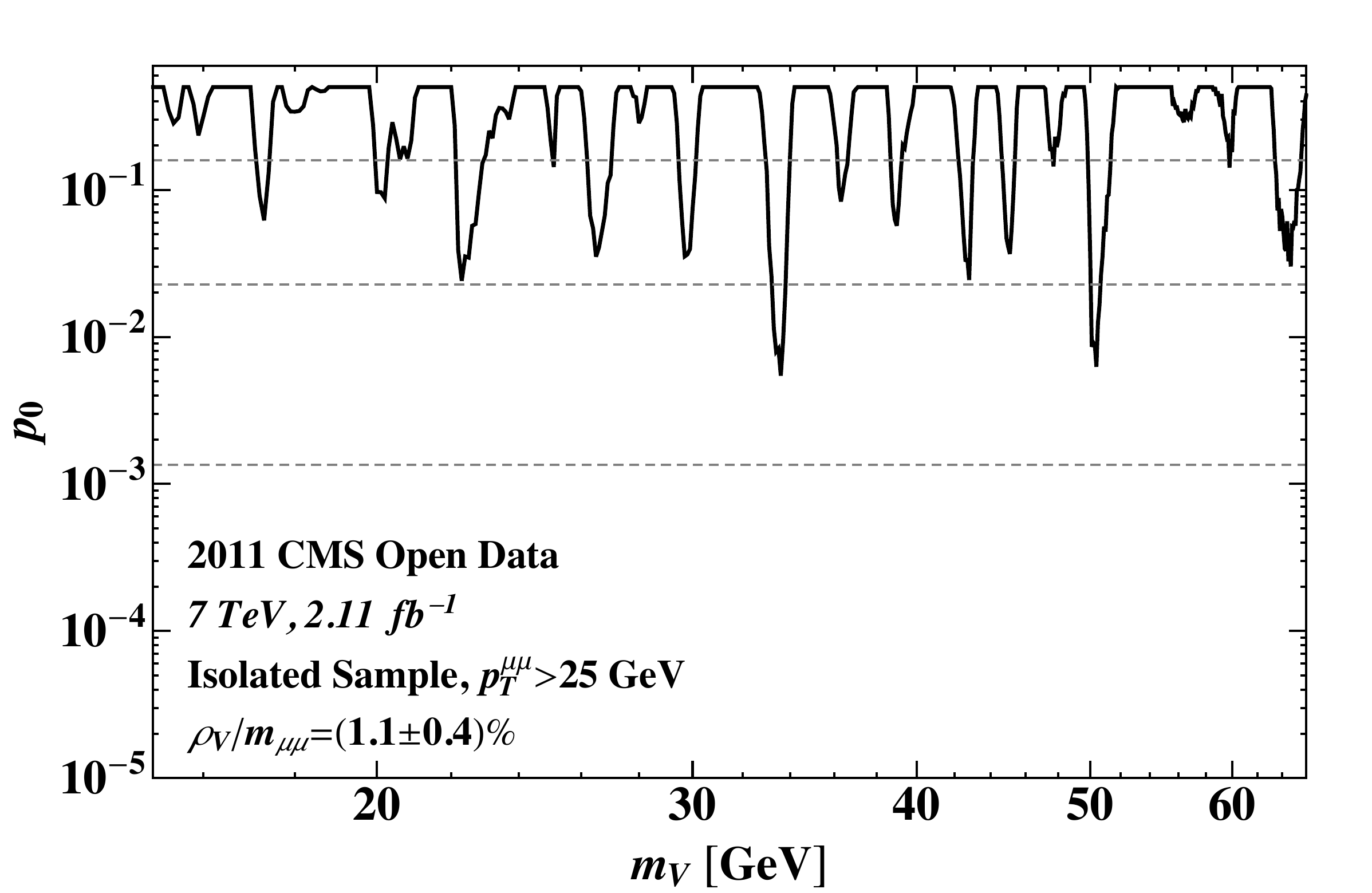} \
\includegraphics[width=\columnwidth]{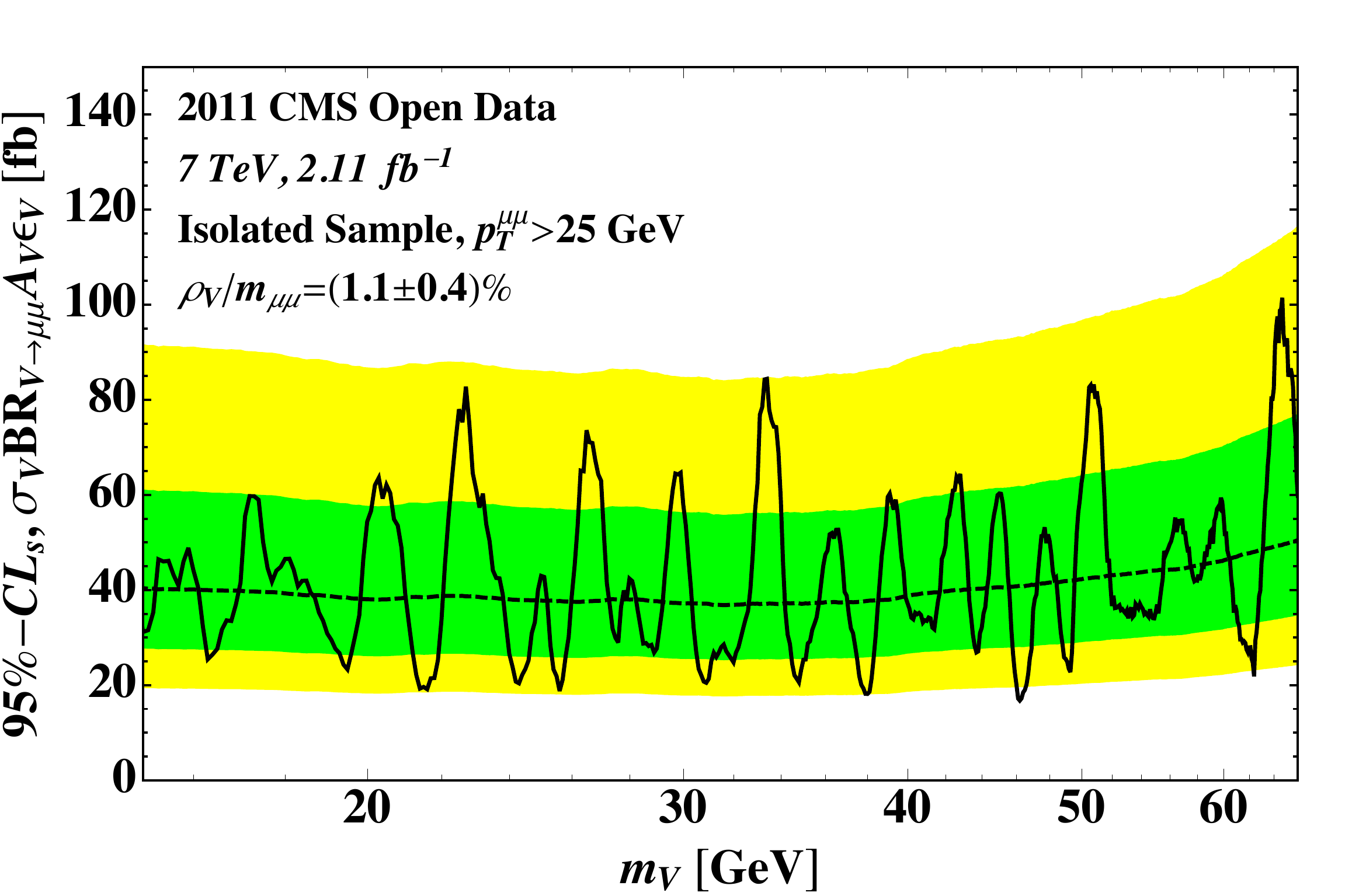} \\
\includegraphics[width=\columnwidth]{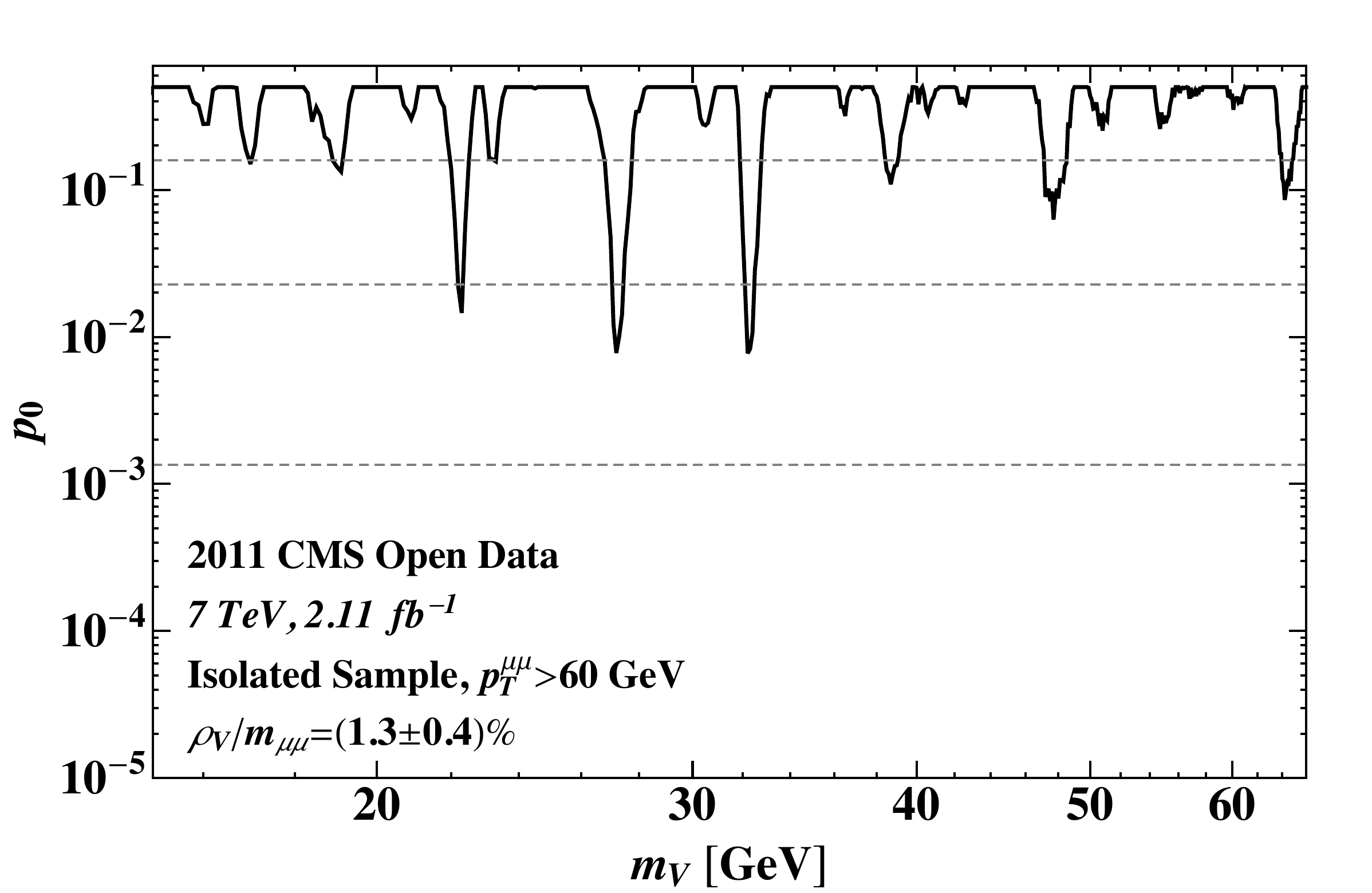} \
\includegraphics[width=\columnwidth]{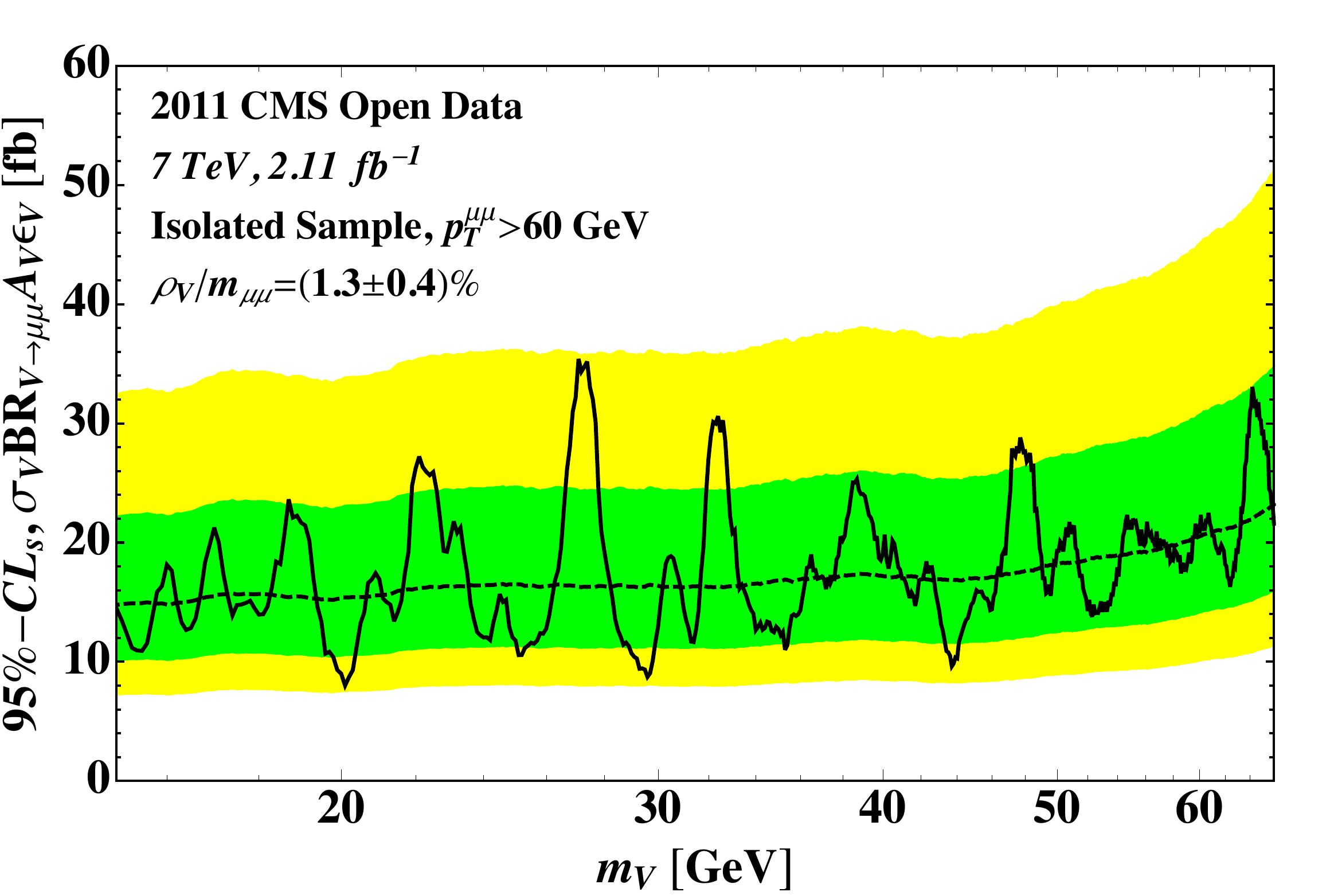} \\
\end{center}
\vskip -0.20in  
\caption{Resonance search for the isolated sample, with (left column) the $p$-value for rejecting the background-only hypothesis as a function of $m_{\mu\mu}$, and (right column) the 95\% $CL_s$ bound, as a function of $m_{\mu\mu}$, on the quantity $\sigma_V \, \BR (V\to\mu^+\mu^-) \, A_V \, \epsilon^V$ defined in  \Eq{eq:limit_iso}, with the expected bound and its 1$\sigma$ (2$\sigma$) bands shown in green (yellow). 
Here, $\sigma_V \equiv \sigma(pp\to V+X)$ is the total $V$ cross section, $A_V$ is the acceptance including the cut on $\ptmm$, and $\epsilon^V \equiv \epsilon^V_{\rm tr} \, \epsilon^V_{\rm iso}$ is the combined trigger/reconstruction and isolation efficiency.
Shown are results with (top row) no $\ptmm$ cut, (middle row) $\ptmm>25$ GeV, and (bottom row) $\ptmm>60$ GeV.
We assume a luminosity of 2.11 fb\inv.
}  
\label{fig:Prf2aOSisoPlots}
\end{figure*}

We now show limits on $V \to \mu^+ \mu^-$ production from our \nameofsearch dimuon search.
Results for the isolated sample are shown in \Fig{fig:Prf2aOSisoPlots}, for the three $\ptmm$ cuts.
Due to trigger-related effects, we show results only for $m_V>33$ GeV  for the inclusive $\ptmm>0$ subsample; below 20 GeV, the $\ptmm>0$ and $\ptmm>25$ GeV samples are very similar and thus redundant, while between 20 and 35 GeV, the rapid variation of the data makes our methods unreliable.
Results could be obtained if the trigger threshold shape could be precisely predicted {\it a priori}, but this is not possible for us, especially for the prompt sample.

%
\begin{figure*}[t]  
\begin{center}  
\leavevmode
\vskip 0.0in
\includegraphics[width=\columnwidth]{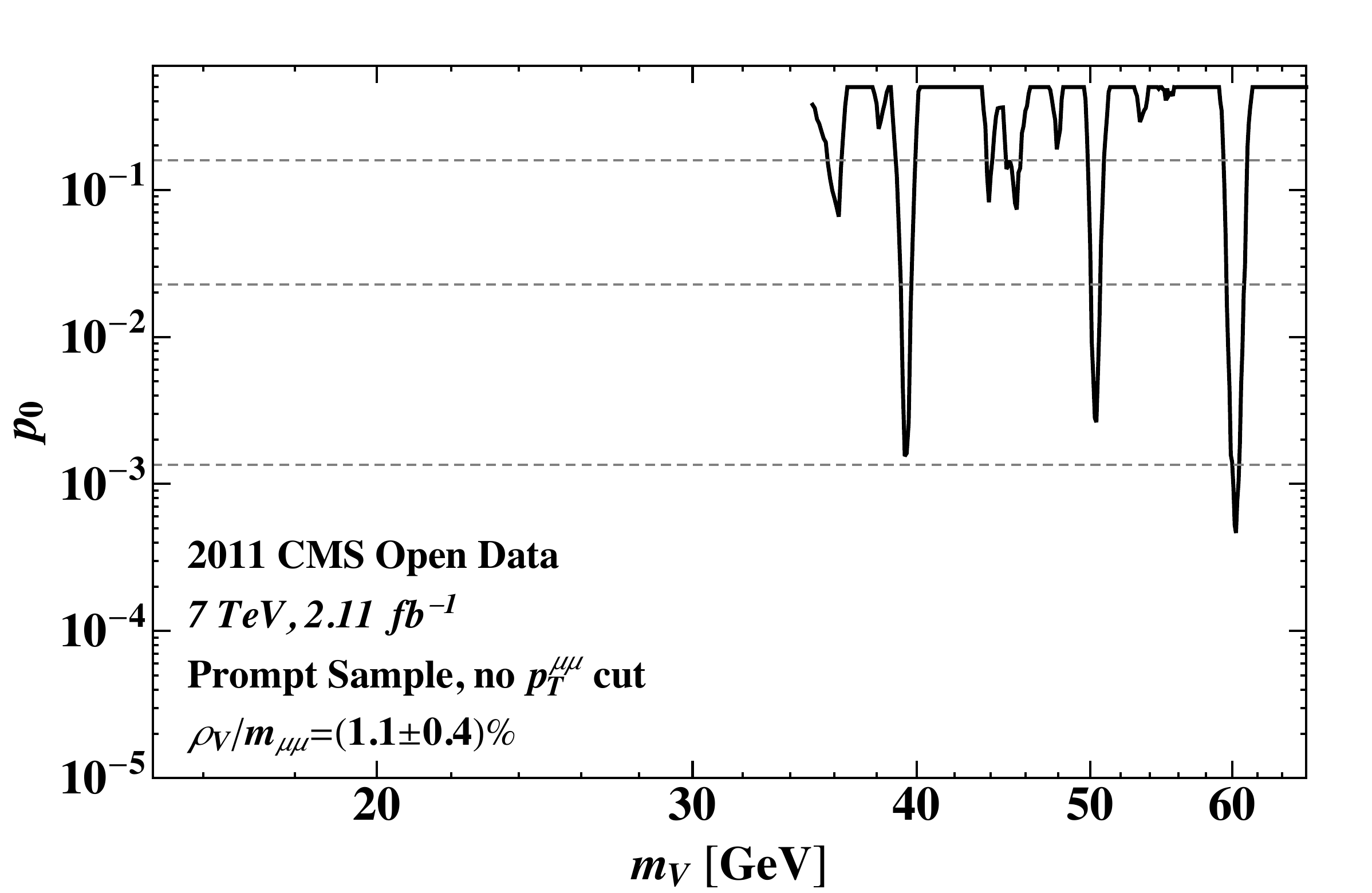} \
\includegraphics[width=\columnwidth]{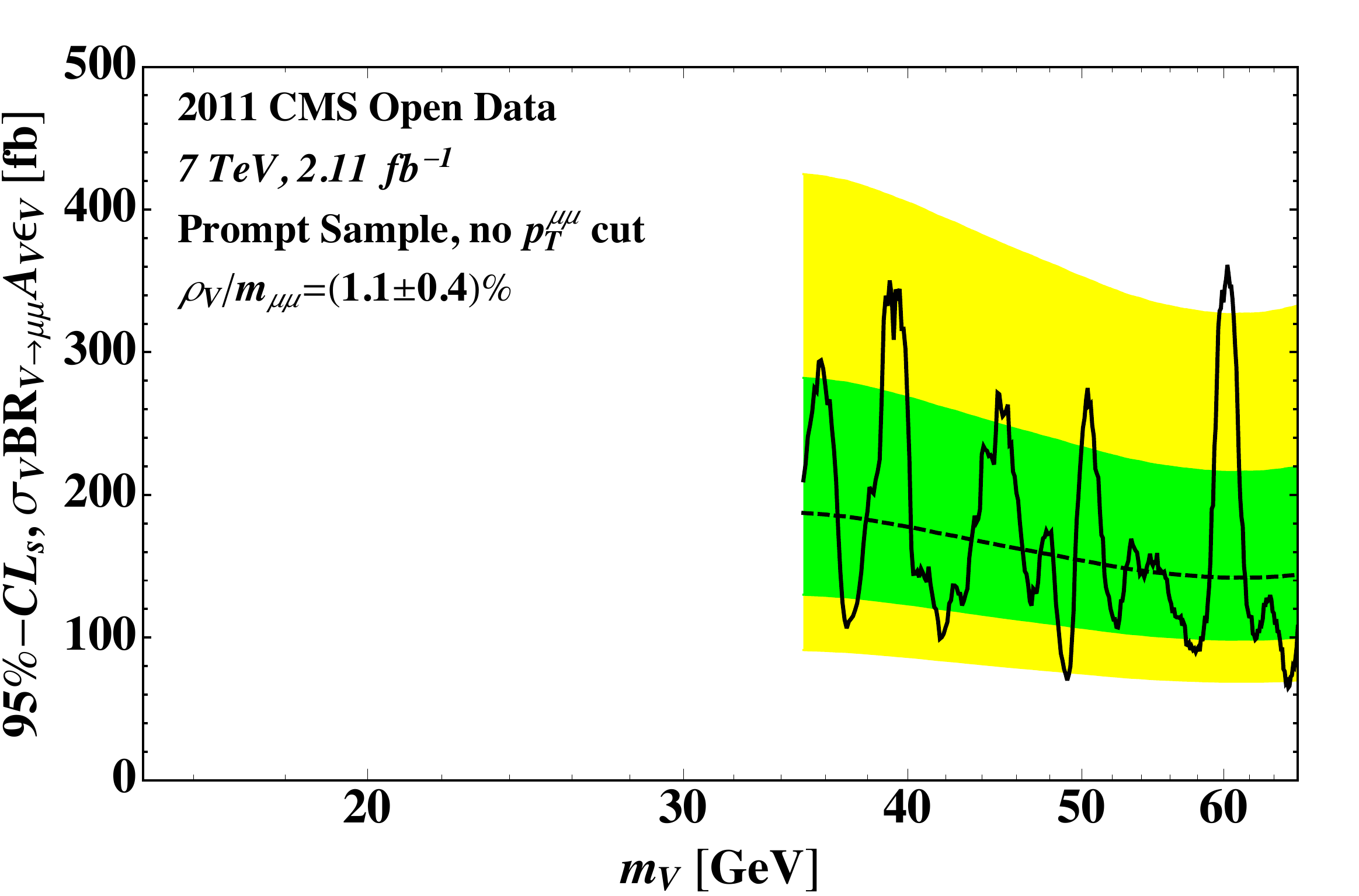} \\
\includegraphics[width=\columnwidth]{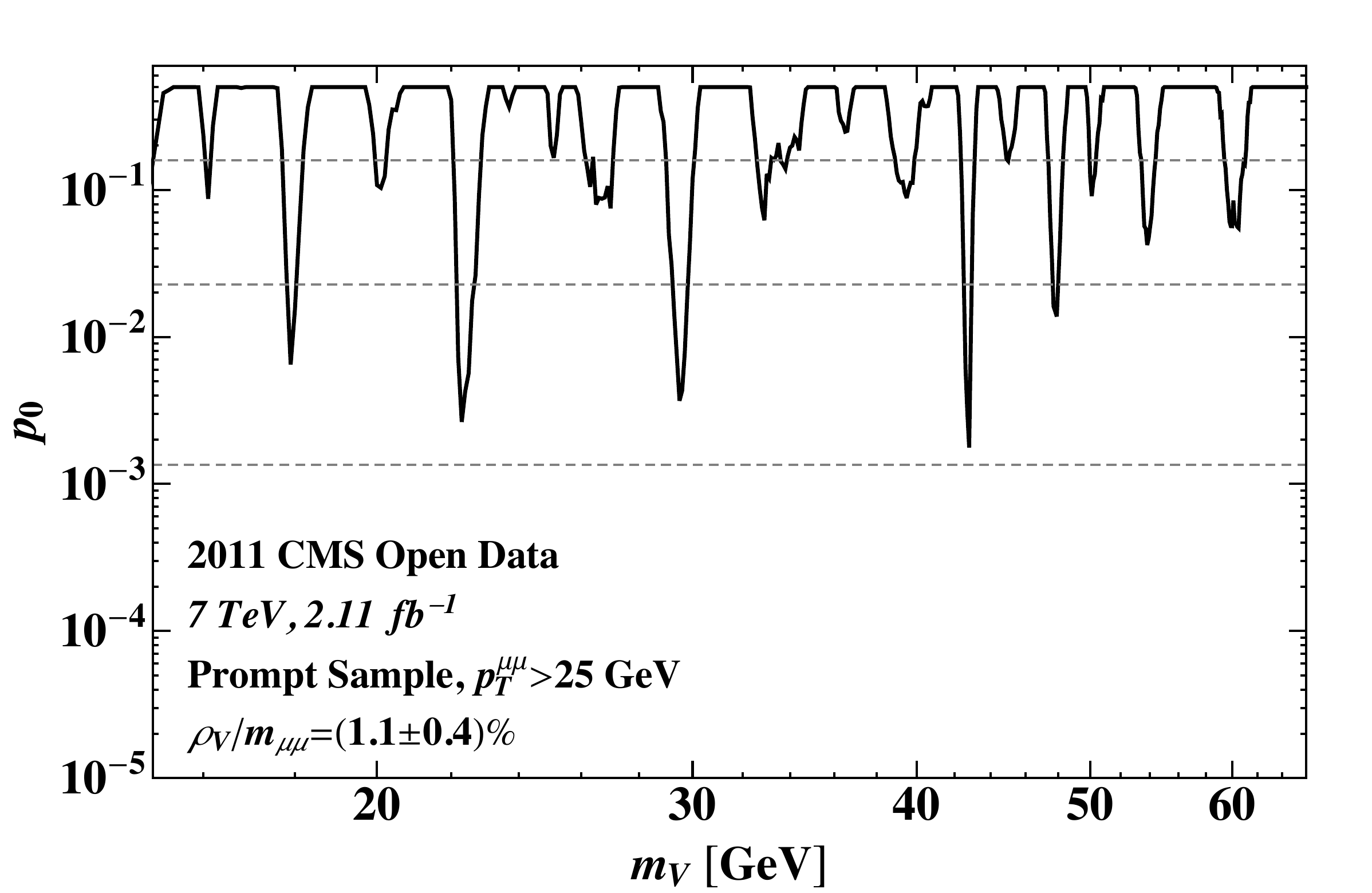} \
\includegraphics[width=\columnwidth]{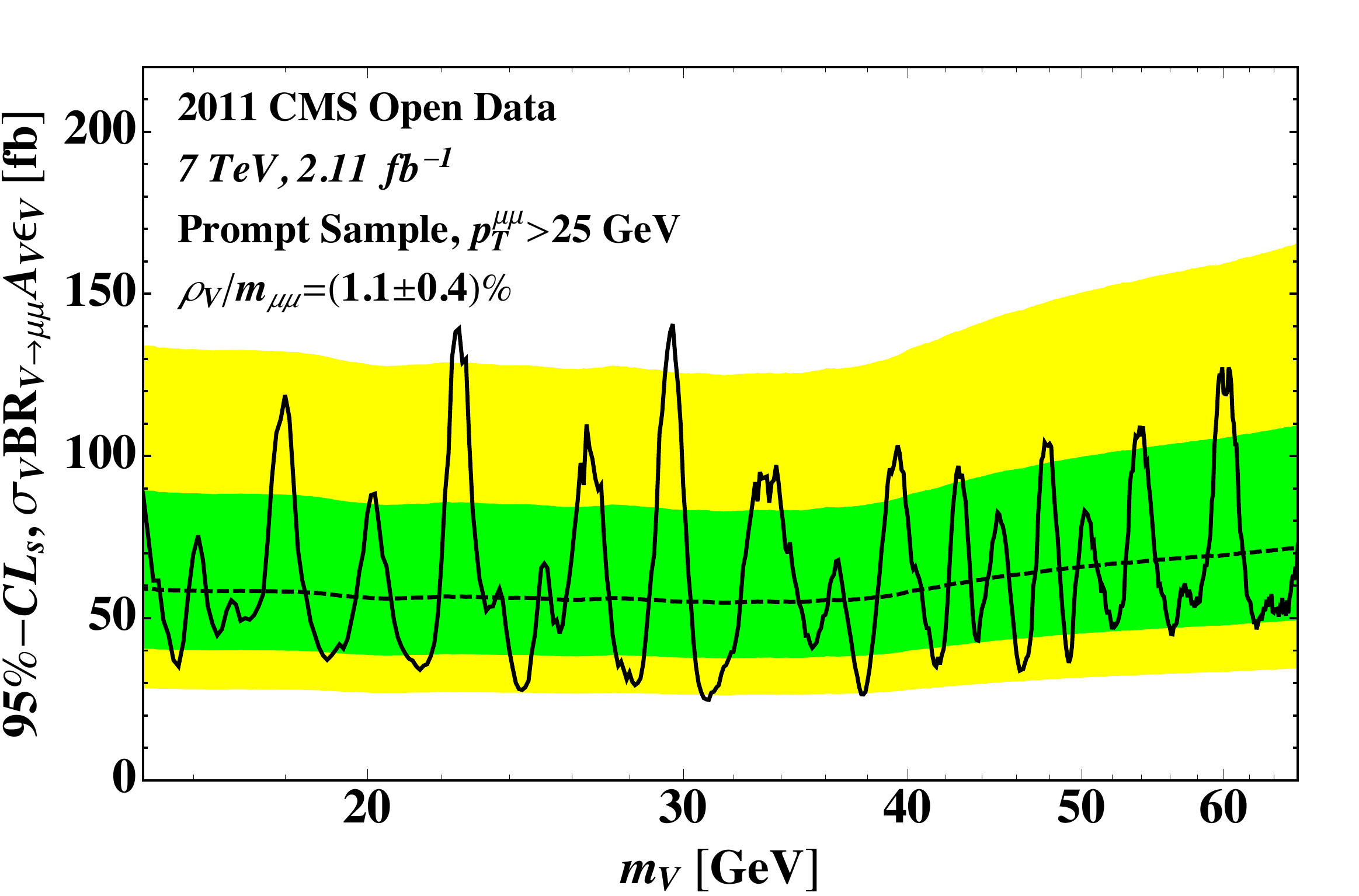} \\
\includegraphics[width=\columnwidth]{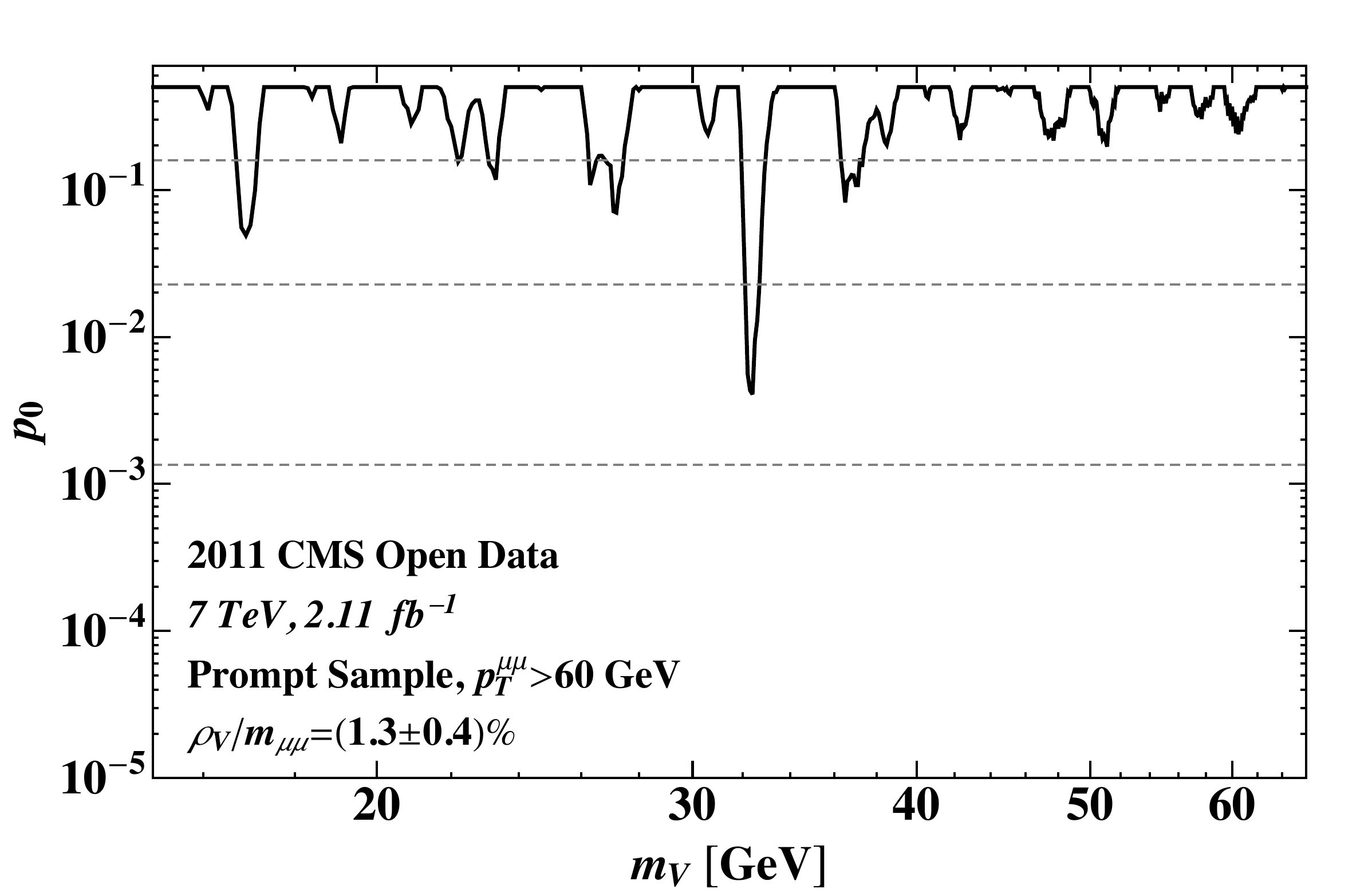} \
\includegraphics[width=\columnwidth]{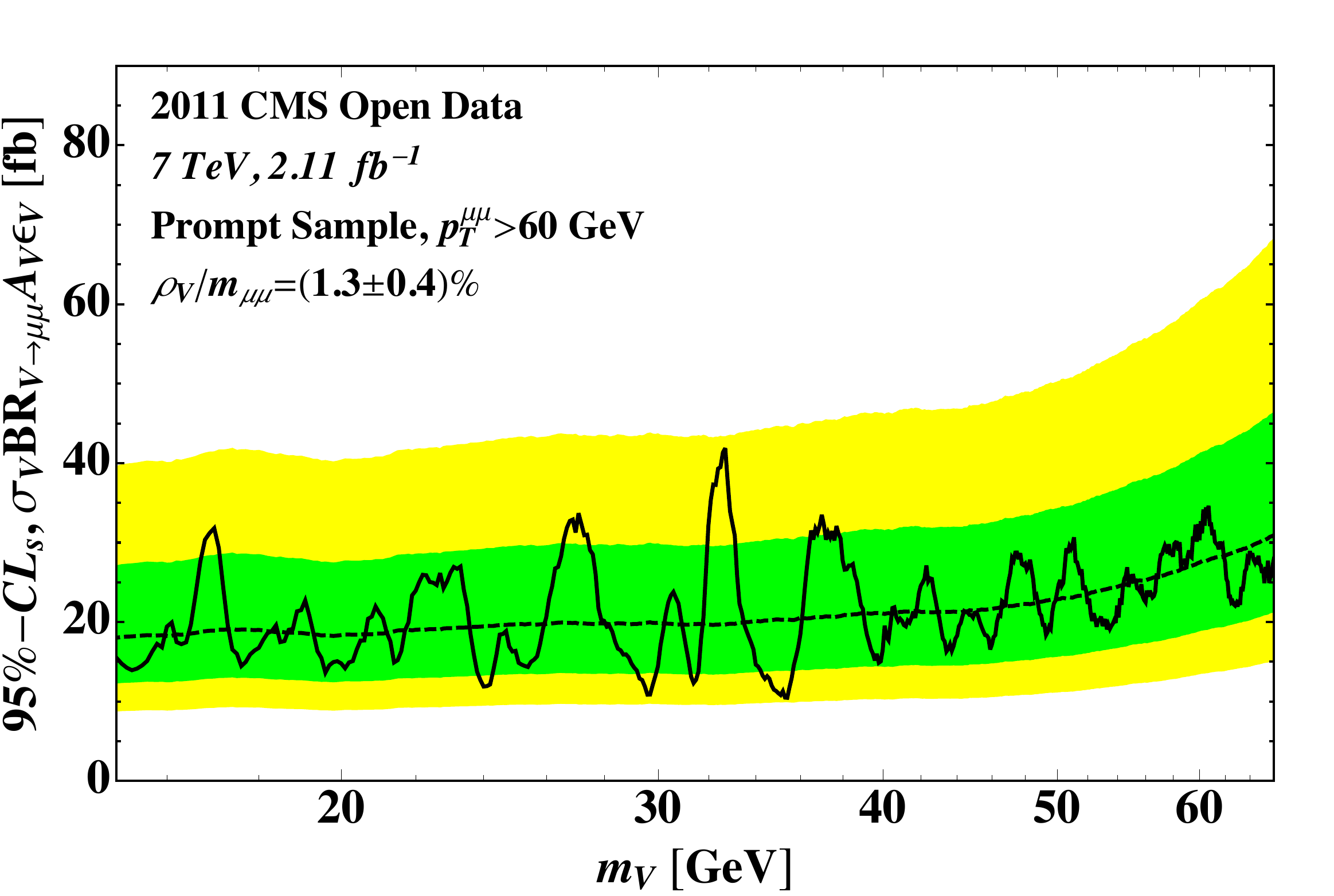} \\
\end{center}
\vskip -0.20in  
\caption{As in \Fig{fig:Prf2aOSisoPlots}, but for the prompt sample.
Bounds are on the quantity $\sigma_V \, \BR (V\to\mu^+\mu^-) \, A_V \, \epsilon^V$ defined in  \Eq{eq:limit_prompt}, now with $\epsilon^V \equiv \epsilon^V_{\rm tr}$.
As described in the main text (see \Tab{tab:SysUnc}) we make a uniform upward correction of 3\% to account for small signal losses from to the tight IP cut.}  
\label{fig:Prf2aOSprPlots}
\end{figure*}

The left column in \Fig{fig:Prf2aOSisoPlots} shows the $p$-values as a function of $m_{V}$, and the right column shows the observed and expected 95\% $CL_s$ upper bounds on the quantity
\begin{align}
\label{eq:limit_iso}
	\sigma(pp\to V+X) \, \BR (V\to\mu^+\mu^-) \, A_V \, \epsilon^V_{\rm tr} \, \epsilon^V_{\rm iso} \ , 
\end{align}
namely, the product of the $V$ production cross section, its branching fraction to muons, the acceptance for $V$ events to pass our cuts, the combined dimuon trigger/reconstruction efficiency for muons in these events, and the corresponding dimuon isolation efficiency.

Similar results for the prompt sample are shown in \Fig{fig:Prf2aOSprPlots}.
Since there is no need to account for an isolation efficiency, our bound is on 
\begin{align}
\label{eq:limit_prompt}
	\sigma(pp\to V+X) \, \BR (V\to\mu^+\mu^-) \, A_V \, \epsilon^V_{\rm tr} \ . 
\end{align}
Note that we have explicitly corrected for the IP cut efficiency; see \Tab{tab:SysUnc}.

\subsection{Use of the Results}
\label{subsec:use_results}

To use the results of \Figs{fig:Prf2aOSisoPlots}{fig:Prf2aOSprPlots} in a model-specific search, one must generate a signal and compute its acceptance and efficiencies, and then combine that with our limits to obtain a bound on the signal cross section times branching ratio.
For this reason, \Tab{tab:SysUnc} does not include any uncertainties on the acceptance $A_V$ or the efficiencies $\epsilon^V$, since these depend on the specific model that one wants to constrain.
The degree of detail with which this must be done depends on the goals of the user.
In many applications, knowing limits to within a factor of 2 is sufficient, and it is rare that knowing them better than 10\% is both necessary and feasible.
Indeed, signal generation is often done at tree level, or at best at one loop, meaning that substantial uncertainties are intrinsic to the methodology.

The trigger and reconstruction efficiency $\epsilon^V_{\rm tr}$, while not constant, generically has weak model dependence.
Under many circumstances, unless high precision is needed, it is reasonable to take $\epsilon_{{\rm tr}}=0.85 \pm 0.05$  and combine this uncertainty with the comparable or larger uncertainties on the signal generator. 
A key exception is if the typical $V$ has a large transverse boost with $p_T^V/m_V \gtrsim 4$, in which case the muons can often be so collimated that the muon trigger system may fail to detect both muons.%
\footnote{This effect, and a corresponding precipitous loss in efficiency in the forward region, can be seen clearly in the $\eta$ distribution of the $J/\psi$ in 
the CMS11a sample.}
This situation requires a dedicated study of $\epsilon_{{\rm tr}}$.%

By contrast, the isolation efficiency $\epsilon^V_{\rm iso}$ and the acceptance $A_V$ can depend strongly on the specific signal model and its parameters; see \Sec{sec:app}.
Fortunately, acceptance is very similar at generator level and detector level.
For isolation, which we have studied using a combination of CMS MC and CMS data, the situation is more complex.  
If the generator-level dimuon isolation efficiency is low, below 60\%--70\%, the prompt sample should be used instead of the isolated sample, and  $\epsilon^V_{\rm iso}$ is not needed.  %
If it is high ($>85\%$) at generator level, then the absolute difference between generator- and detector-level efficiencies is typically less than 10\% and so an uncertainty of this order may be taken.
In the region between, the differences between generator and detector level must be studied with more care. 
However, for the limits with a $\ptmm$ cut of 25 or 60 GeV, a detector-level isolation efficiency of $\epsilon^V_{\rm iso}=60\%$--$80\%$ makes the sensitivities of the prompt and isolated samples comparable.
The user can then choose whether to use the prompt samples, at the cost of slightly lower but more certain sensitivity, or to study the isolation with more precision so as to benefit from the slightly higher sensitivity of the isolated samples.

A user requiring higher precision will need to estimate $\epsilon_{{\rm iso}}$ and $\epsilon_{{\rm tr}}$, and their uncertainties, as we have done in our $Z$ study above, using information from CMS MC and CMS data, as well as data/MC comparison studies such as in \Ref{Chatrchyan:2012xi}.
Details of how we performed these estimates will be given in future work.%
\footnote{Specifically, since the $p_T$ resolution, the trigger/reconstruction efficiency, and the conversion factors from generator-level to detector-level isolation efficiency are dominantly a function of the single muon $\eta$ and $p_T$, we may try in the future to release this information in the same format as \Ref{Aaij:2018xpt} to allow for easier recasting of our bounds.}
The precision user will also need to account for uncertainties on the acceptance $A_V$, and possible important corrections and uncertainties due to the muon $p_T$ resolution and scale factor.
Finally, the user must estimate the appropriate signal line shape and resolution to confirm it is within the uncertainties of our assumptions in \Sec{subsec:lineshape}, or if not, must correct for it, replacing our line shape with one appropriate to another model.
However, the precision user should also consider that there are small residual uncertainties in the choice of window and fitting function in \Sec{subsec:limit_procedure}, and there is no agreed-upon procedure for quantifying such uncertainties in the literature.

\subsection{Interpretation of the Limits}

Let us now examine the results of \Figs{fig:Prf2aOSisoPlots}{fig:Prf2aOSprPlots}, keeping in mind that the prompt and isolated samples overlap (as do the samples with different $\ptmm$ cuts) and are therefore correlated.
For $m_{\mu\mu}\sim 35$--$45$ GeV, {\it i.e.}\ where the trigger is efficient, the cut $\ptmm> 25$\,(60) GeV gives expected bounds, relative to the sample with no cut, that are smaller by a factor of $\sim 3\,(6)$ for the isolated sample and a factor of $\sim 3\,(9)$ for the prompt sample.
For $m_{\mu\mu}$ well below $35$ GeV, the $\ptmm> 60$ GeV cut gives expected bounds smaller than the $\ptmm> 25$ sample by slightly less (more) than a factor of $3$ for the isolated (prompt) sample.
More specifically, in the isolated sample, our expected bounds are in the range of 40\,(15) fb for $\ptmm> 25$\,(60)~GeV, and correspondingly 60 (20) fb for the prompt sample.

The most significant excursions from expectation in the $p$-value plots are for the inclusive prompt sample, in the 2$\sigma$--3$\sigma$ range.
However, an estimate of the global $p$-value for this plot, following the methods of \Ref{Gross:2010qma}, gives 0.032, slightly below 2$\sigma$ significance.
(This result is obtained by counting up-crossings at a baseline significance-squared of $u_0=0.5$; changing this to $0.25$ or $1$ leaves the answer nearly unchanged.)
The  global significance of the other plots is  below 1$\sigma$, including the prompt $\ptmm>25$ GeV subsample whose largest local excess (discussed further below) is nearly 3$\sigma$.

One excess, at 29.5 GeV in the prompt sample with $\ptmm>25$ GeV, merits a mention since it lies in a region that is already of some interest~\cite{Heister:2016stz,Sirunyan:2018wim} (see \Refs{Godunov:2018qsu,vanBeveren:2018hnp} for follow-up phenomenological studies).
At this mass value, the background is rejected at $2.7\sigma$ local significance.
Most likely this is a statistical fluctuation; two spikes of comparable size appear elsewhere in the same plot, and another appears at 32.5 GeV for $\ptmm>60$ GeV.
However, let us briefly consider whether this excess could possibly reflect a signal.
No corresponding spike is present for the sample with $\ptmm>60$ GeV, but this does not by itself argue against a signal; we will see examples of signals with this behavior in \Sec{sec:app} ({\it e.g.}\ the dotted red curve in \Fig{fig:pt_signals_vs_bkgd}).  
Also note that this excess may not be inconsistent with the results from CMS at this mass range~\cite{Sirunyan:2018wim}, because even though CMS has larger samples from both Run I and Run II, their analysis imposes different cuts (requiring a $b$ tag and a central jet veto), which would have very low acceptance for certain signals to which we would be sensitive.
For any particular signal, a detailed recasting of the CMS results would be needed, beyond our scope here.\footnote{Our analysis is insensitive to the specific excess in $Z$ decays observed in \Ref{Heister:2016stz}, despite hundreds of expected events in CMS11a.
As shown in Appendix B of \Ref{Heister:2016stz}, the typical $|\vec p\,|^{\mu\mu}$ of the excess is low in the $Z$ frame.
In the CMS11a data, then, our $\ptmm$ cut has very low acceptance, unless a second production mechanism at the LHC creates additional dimuons at higher $\ptmm$.}

The most dramatic $p$-value spike in the $\ptmm>25$ GeV plot, at 42.7 GeV, has been unrealistically enhanced as a result of the large uncertainty in the resolution $\rho_V$ (adopted from the CMS recommendation; see \Sec{subsec:lineshape} above).  
This is reflected in the extreme narrowness of the spike and lack of a similarly large excess in the limit plot at that mass. 
This effect can occur when an excess in the data has a width smaller than the central value $\langle\rho_V\rangle$, in which case the fit to a narrow signal may be excellent, resulting in a very small $p$-value.
On the other hand, a narrower signal faces smaller backgrounds, so the observed limit (for a fixed $p$-value) is lower than would be expected for a significant signal with width $\langle\rho_V\rangle$.
The excursion of the observed limit above the expected limit is therefore relatively small.
A reduced uncertainty on $\rho_V$ on the low side, as our $J/\psi$ studies suggest would be appropriate, would make the $p$-values at such locations less significant, with little effect on the observed limits at those masses.
We have confirmed this by profiling over the mass resolution using ${}^{+0.4\%}_{-0.2\%}$ instead of the nominal $\pm0.4\%$; the most dramatic effect  is to reduce the significance of the $p$-value peak at 42.7 GeV by $0.5\, \sigma$.
Little or no effect on other $p$-values or on the limits is seen in this or other samples.
Thus, at locations with significant $p$-value spikes but a much less significant excess in the limit plot, some caution is advisable.

We additionally caution that small changes in our fitting method can lead to shifts in the local significance of excesses of order $0.5\sigma$. 
(Changes to the expected and observed limits are smaller.)  
For example, adjusting the fitting window from $35\rho_V$ to $30\rho_V$ or $40\rho_V$ is sufficient to see effects of this size, as is using the cubic model instead of the quintic one.

One can only say, therefore, that the data show no clearly significant excesses.  
What is more essential, however, is that  application of our methods to Run II data would lead to limits an order of magnitude stronger.
Such an analysis would immediately reveal or exclude any particle hypothetically responsible for any of the excesses in our plots.

%
\begin{figure}  
\begin{center}  
\includegraphics[width=0.96\columnwidth]{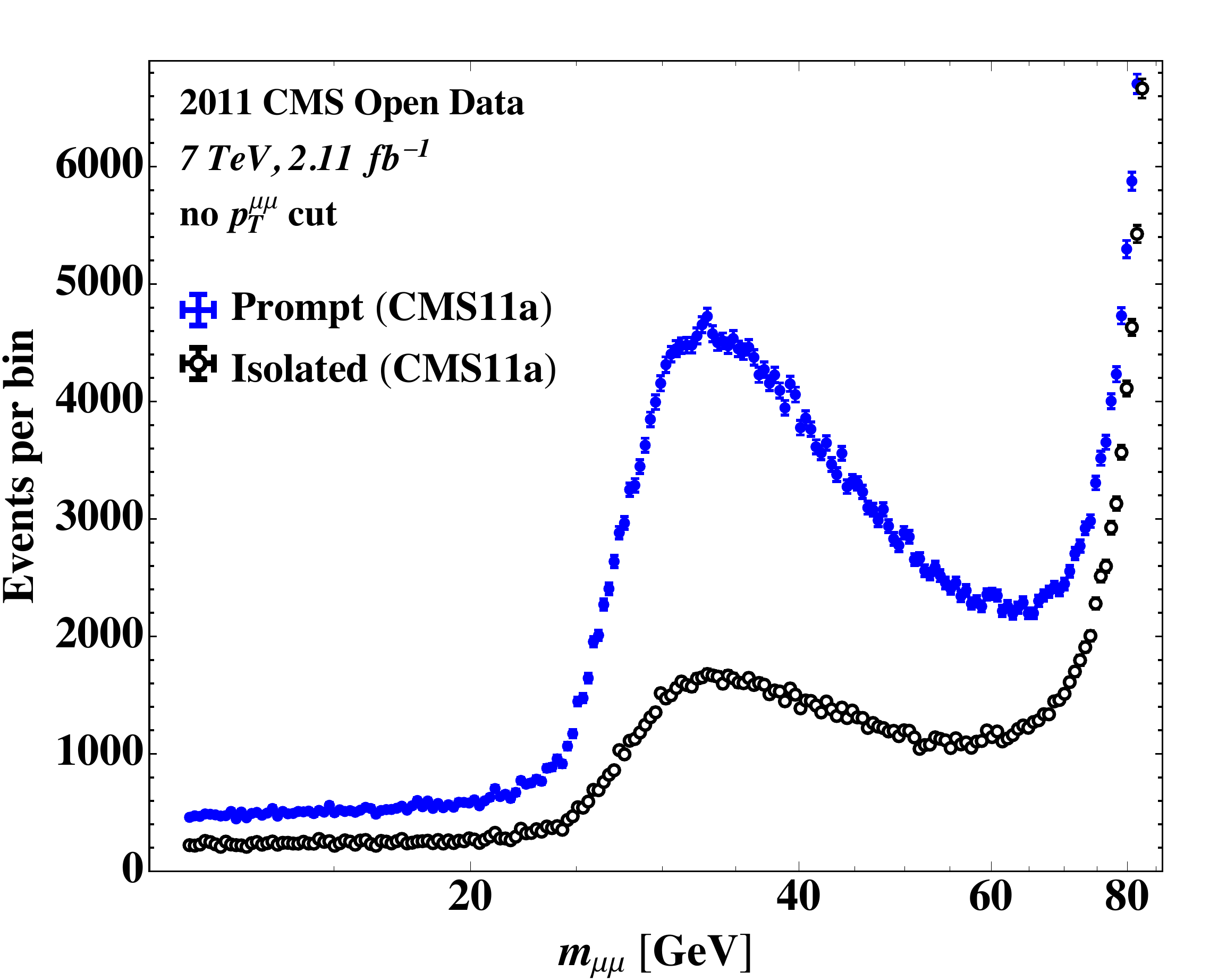}
\includegraphics[width=0.96\columnwidth]{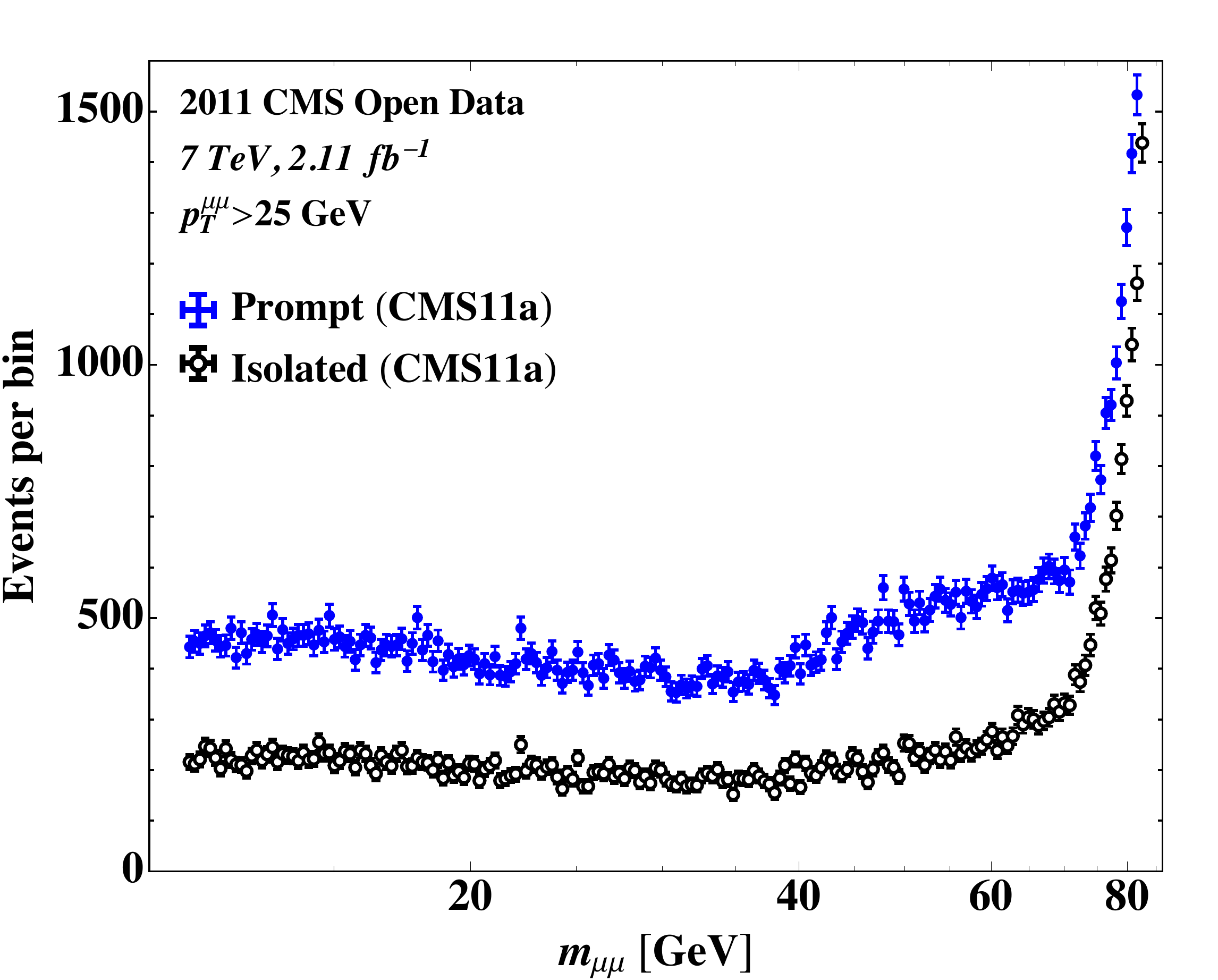}
\includegraphics[width=0.96\columnwidth]{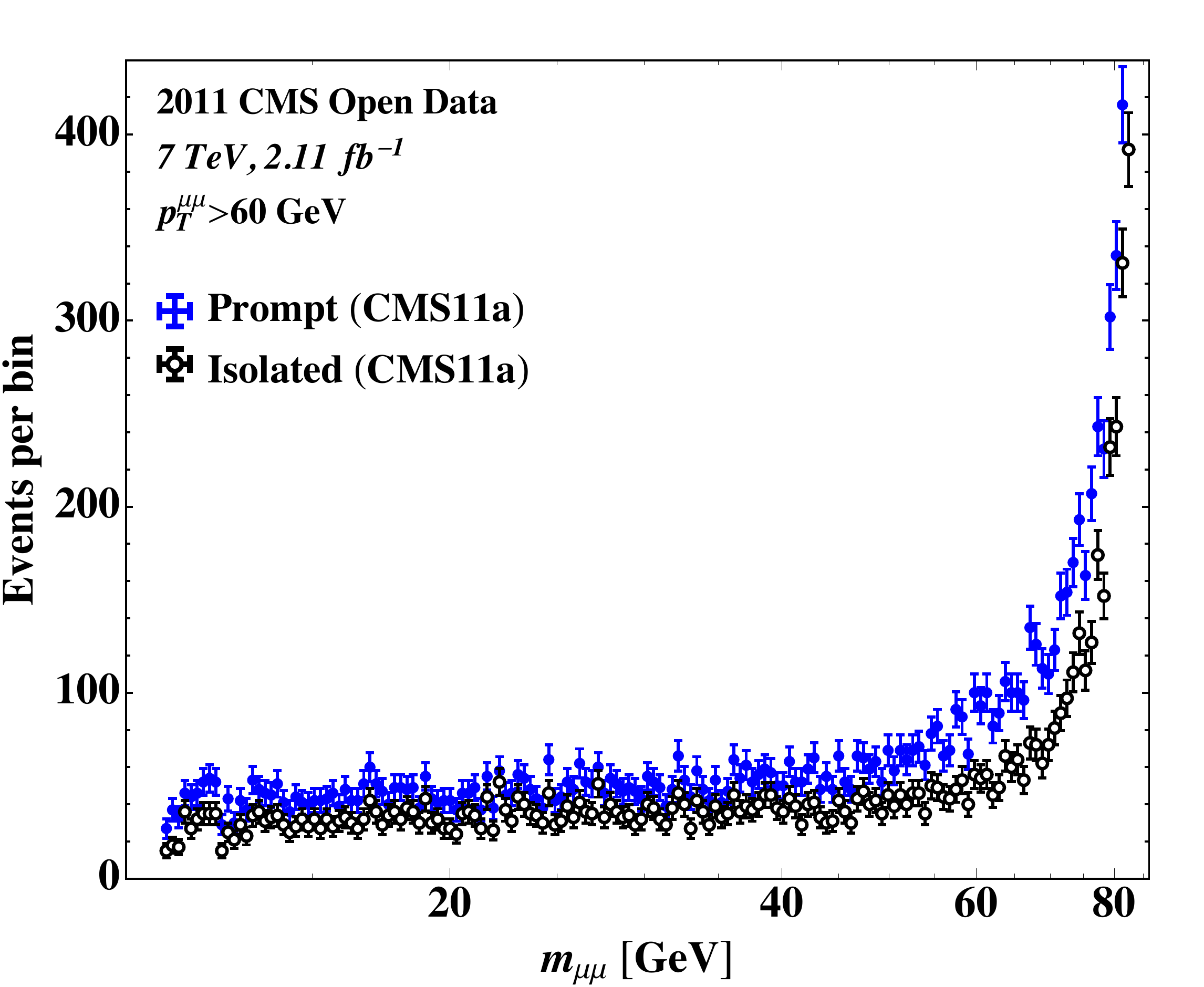} 
\end{center}
\caption{The dimuon mass spectrum of the prompt (upper points, blue) and isolated (lower points, black) CMS11a samples, after all other quality and kinematic cuts. 
Shown are distributions with (top) no $\ptmm$ cut, (middle) $\ptmm>25$ GeV, and (right) $\ptmm>60$ GeV.  Bins are chosen equal to the resolution appropriate to the plot (1.1\% for the upper plots, 1.3\% for the lower plot).
} 
\label{fig:CompareP25PrVsIso}
\end{figure}

As a further check, we show the dimuon spectrum with $\ptmm> \{0 \GeV,25 \GeV,60 \GeV\}$ in \Fig{fig:CompareP25PrVsIso}.
 For the $\ptmm>25$ GeV samples, the number of events is such that all the 2$\sigma$ excursions can be seen by eye, giving a useful cross-check on our results.
This figure also illustrates our earlier remark that, while there is virtually no QCD background in the isolated sample, the DY and QCD backgrounds are of similar size in the prompt sample, with QCD falling faster  with $p_T$ than DY.

Let us note, finally, that only technical issues deter us from applying stronger $\ptmm$ cuts, or from searching at higher or lower masses.
At higher masses and/or with higher $\ptmm$ cuts, the event counts become very low and our fitting procedure requires more care; the strategy of \Ref{Williams:2015xfa} may be helpful in this context. 
At lower masses and/or with higher $\ptmm$ cuts, muons become increasingly collimated.
As mentioned above, excessive collimation causes the muon trigger system to become inefficient at separating the two muons, especially at high $|\eta|$.
A more careful study of trigger and reconstruction efficiencies (or use of the much larger single muon stream) would be required.
We do not address these issues here, but nothing should prevent the LHC experimental collaborations from extending a \nameofsearch dimuon search strategy into these more extreme kinematic regions.

\section{Implications for Benchmark Scenarios}
\label{sec:app}

In this section, we briefly consider the implications of our bounds for benchmark signals.  
As discussed in \Sec{subsec:use_results}, full application of the bounds requires detailed discussion of how to obtain the various efficiencies for a particular model, which will be presented in future work.
Here, we simply demonstrate that simple models exist in which $A_V$ remains large with our $\ptmm$ cuts (and $\epsilon^V_{\rm tr}$ is unsuppressed).
For these models, which include cases where the $V$ is produced in the decay of a heavier particle,
our \nameofsearch search strategy offers much improved sensitivity, because the trigger/reconstruction efficiency $\epsilon_{\rm tr}$ is mostly independent of the $\ptmm$ cut, and any significant change in isolation efficiency can be addressed through the judicious use of the isolated and prompt samples.
By contrast, as we discuss at the end of this section, our strategy is not aimed at the minimal dark photon models~\cite{Okun:1982xi,Galison:1983pa,Holdom:1985ag,Pospelov:2007mp,ArkaniHamed:2008qn,Bjorken:2009mm}, where $V$ is predominantly produced via kinetic mixing with the photon/$Z$ of strength $\epsilon$.

\subsection{Production of $V$ via Decay}
\label{subsec:viadecay}

\begin{figure}[t]  
\begin{center}  
\includegraphics[width=0.93\columnwidth]{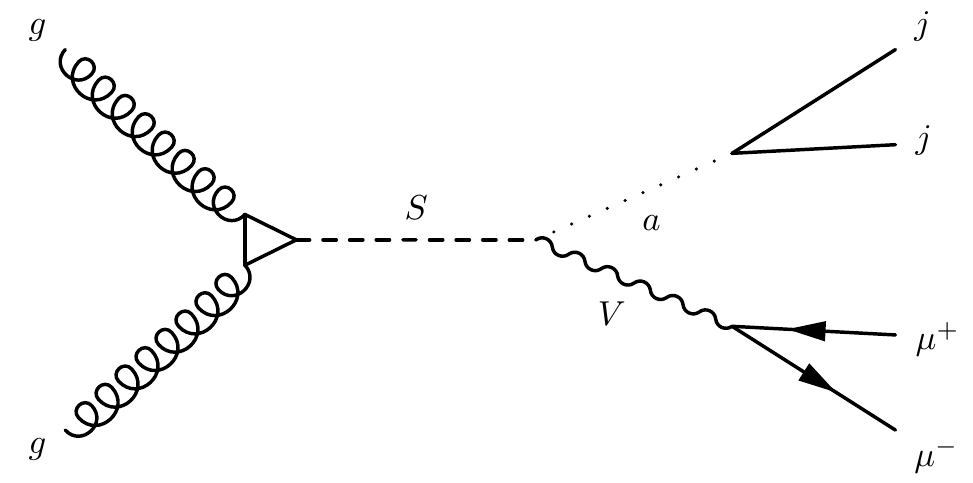}
\includegraphics[width=0.93\columnwidth]{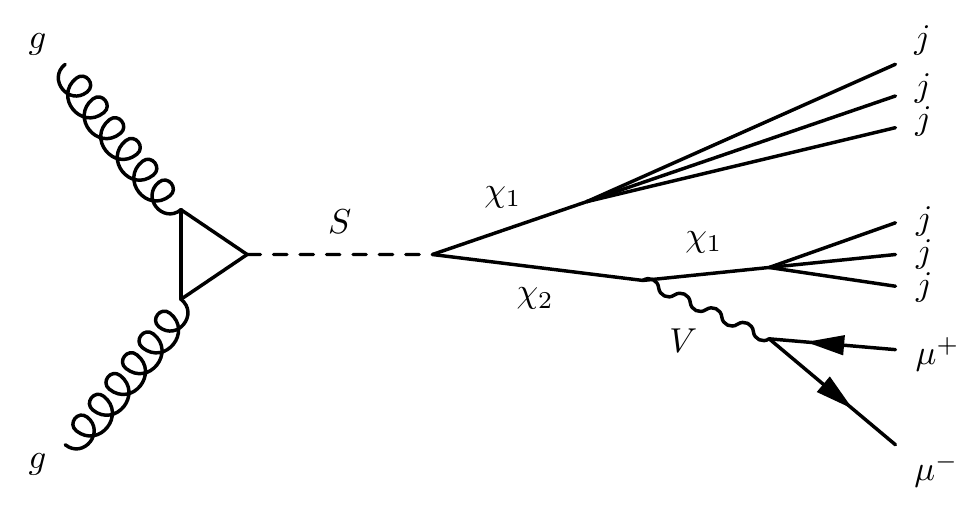} 
\end{center}
\caption{Feynman diagrams for two benchmark models that produce a $V$ boson with substantial transverse momentum.  In some cases, the scalar $S$ might be identified with the SM Higgs boson.}
\label{fig:diagrams}
\end{figure}

\begin{figure}[t]  
\begin{center}  
\leavevmode
\vskip 0.0in
\includegraphics[width=0.93\columnwidth]{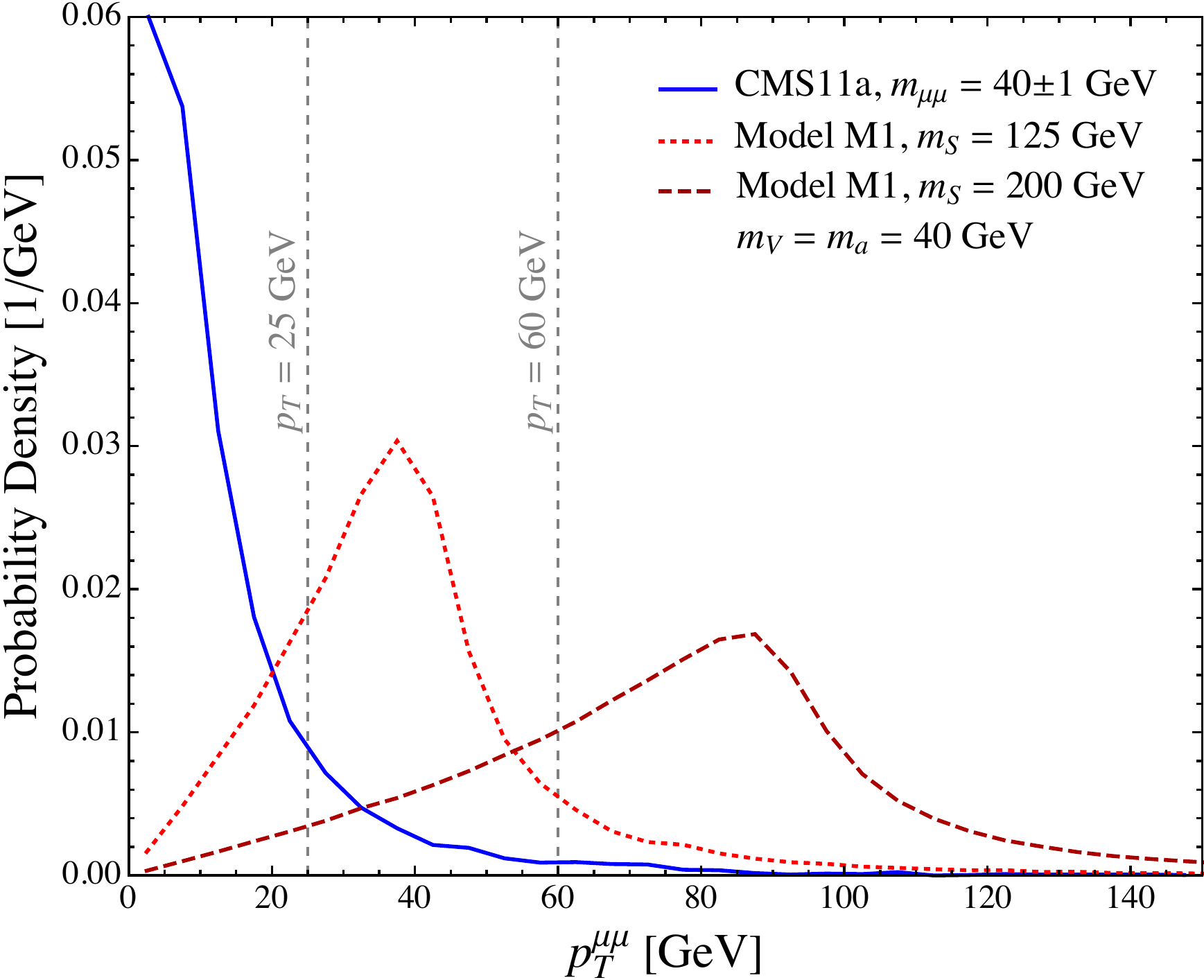}
\end{center}
\vskip -0.20in  
\caption{The $\ptmm$ distribution (normalized to unity) for the events in CMS11a with $m_{\mu\mu}$ between 39 and 41 GeV (solid blue), compared to the corresponding distribution for $V$ at two different parameter points for M1: $m_V=m_a=40$ GeV and $m_S=125$ GeV (dotted red) and $m_S=200$ GeV (dashed dark red).}
\label{fig:pt_signals_vs_bkgd}
\end{figure}

In models where the $V$ is produced predominantly in the decay of a heavier particle, our $\ptmm$ cuts often increase sensitivity.
To see this, consider the two simple theoretical models shown in \Fig{fig:diagrams}, which both contain a scalar $S$ (possibly identified with the 125 GeV Higgs $h$) and a vector $V$ that decays to muons:
\begin{itemize}
\item M1:  $S\to V+a$, where $a$ is a pseudoscalar dominantly decaying to gluon pairs (or perhaps to $b\bar b$); and
\item M2:  $S\to \chi_1\chi_2$, $\chi_2\to \chi_1+V$, $\chi_1\to qqq$, where $\chi_i$ are neutral fermions and the decay of $\chi_1$ is similar to that of an LSP in R-parity-violating supersymmetry. 
\end{itemize}
If $m_V,m_a\ll m_S$ in M1, or if either $m_V+m_{\chi_1} \ll m_{\chi_2}$ or $m_{\chi_2}+m_{\chi_1}\ll m_S$ in M2, the $V$ resonance will have substantial $p_T$ in most events.
In both models, the final state of interest is $\mu^+\mu^-$ plus jets and no missing transverse momentum, for which there are few searches at the LHC.\footnote{One exception is \Ref{Chatrchyan:2012tw}, though that search required the dimuons to reconstruct a $Z$ boson and imposed the equivalent of $S$ to have mass above 500 GeV.}
In \Fig{fig:pt_signals_vs_bkgd} we show the dimuon $p_T$ distribution (normalized to unity) in model M1 for $m_V=m_a=40$ GeV and for two choices of $m_S$, along with the $p_T$ distribution of the background in CMS11a between 39 and 41 GeV.  
The peaking of the signal above a rapidly falling background makes clear why our cuts are effective for models in this class.

%

In model M2, if $m_S>2 m_{\chi_2}$, then $S\to \chi_2\chi_2$ could potentially occur and produce four-lepton events, which are powerfully constrained by multi-lepton searches.  
For any $m_V$, however, there are choices of $m_S$ and $m_{\chi_2}$ where this is kinematically forbidden to occur on shell, while still allowing $S\to \chi_1\chi_2$.
Furthermore, in some models $S\to \chi_2\chi_2$ can be highly suppressed, for example by approximate symmetries or small couplings.  
In any case, our analysis is model independent, so the fact that other searches may rule out some parts of parameter space for particular models does not affect the validity of our results.

\begin{figure*}[t]  
\begin{center}  
\leavevmode
\vskip 0.0in
\includegraphics[width=\columnwidth]{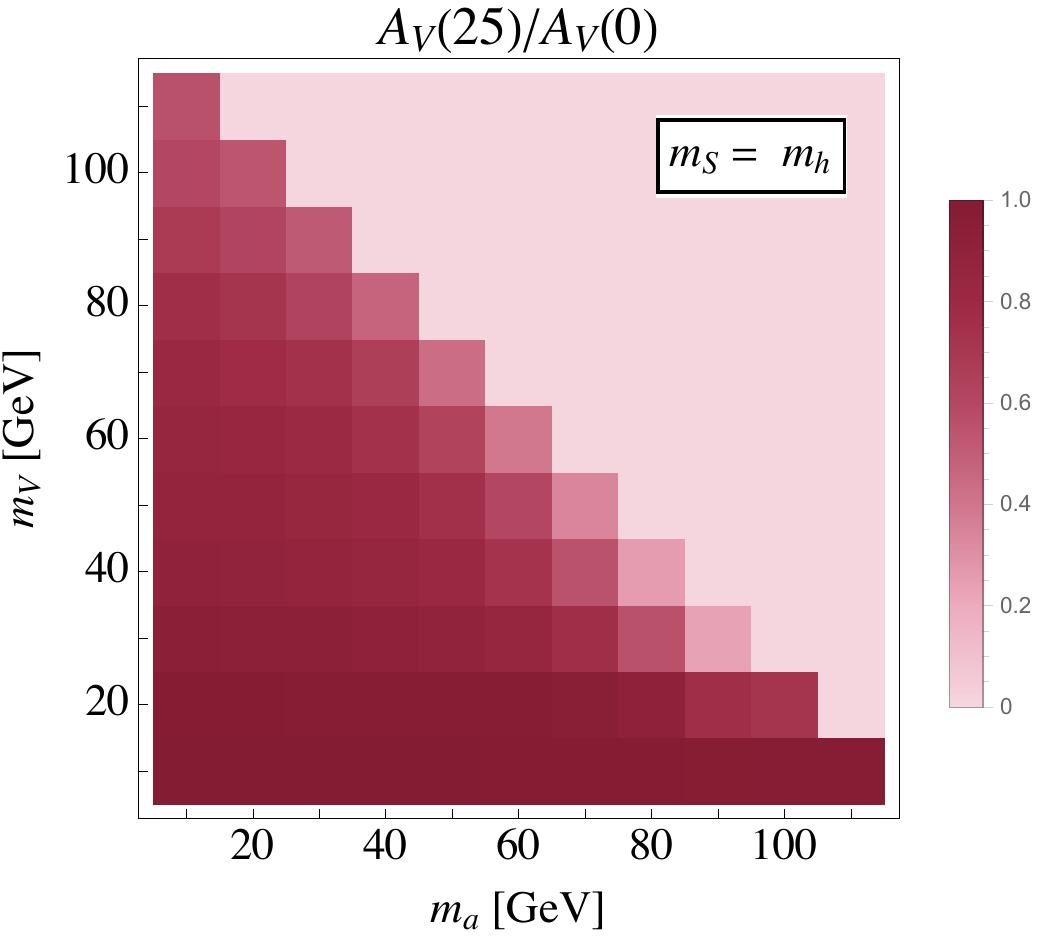} \
\includegraphics[width=\columnwidth]{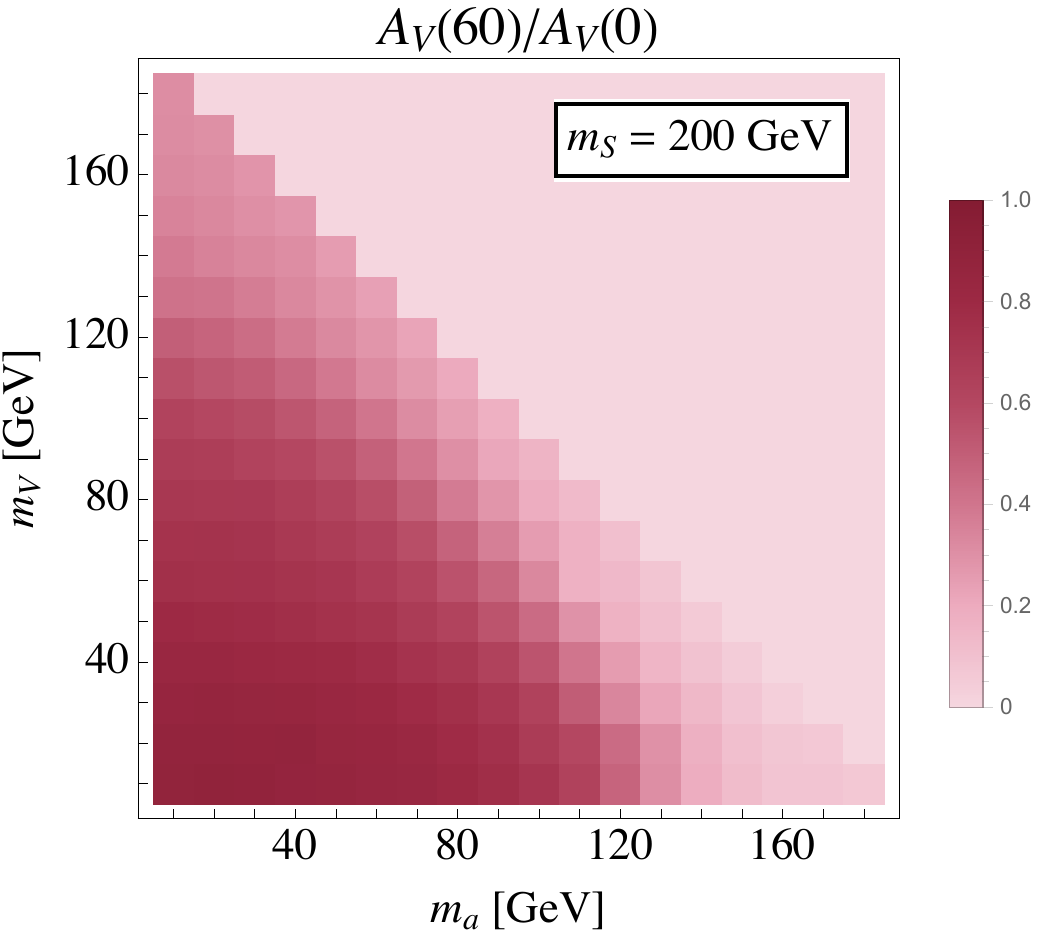} 
\end{center}
\vskip -0.20in  
\caption{After all other cuts, the relative acceptance in model M1 when a $\ptmm$ cut is applied, as a function of $m_V$ and $m_a$.
The relative signal acceptance is often over 50\%, justifying the use of the $p_T$-enhanced search strategy.
Left: For $m_S=125$ GeV, the ratio of the acceptance $A(25)$ for a $\ptmm>25$ GeV cut over the acceptance $A(0)$ with no $\ptmm$ cut.
Right: The same, but for $m_S=200$ GeV, and for the acceptance $A(60)$ for a $\ptmm>60$ GeV cut over $A(0)$.}
\label{fig:M1_125ratio25_200ratio60}
\end{figure*}

For model M1, we expect the isolated sample to yield the best limits, since the decay products of the pseudoscalar  $a$ are unlikely to contaminate the muon isolation cones.
To assess the degree to which the \nameofsearch dimuon strategy improves upon an inclusive search, consider the case that $S$ is identified with the 125 GeV Higgs boson. 
Using \textsc{Pythia}~8.235~\cite{Sjostrand:2014zea}, we estimated the signal acceptance as a function of the $\ptmm$ cut, namely $A_V(\ptmm/\text{GeV})$.
The {\it absolute} signal acceptance for the inclusive search is $A_V(0)\sim 50\%$--80\% for $m_V\geq m_a$.
But the relevant quantity when evaluating the benefits of a $\ptmm$ cut is the {\it relative} acceptance between a \nameofsearch search with, say, $\ptmm>25$ GeV and an inclusive search with no $\ptmm$ cut.
In \Fig{fig:M1_125ratio25_200ratio60} (left), we see that $A_V(25)/A_V(0) \sim 60\%$--100\% when $m_V>m_a$ and $m_V+m_a<100 \GeV$.
(This is not surprising since, for $m_a=m_V<57$ GeV, the $V$ momentum in the $S$ rest frame always exceeds 25 GeV.)
Since our expected bounds for $p_T^{\mu\mu}>25$ GeV and $m_V>33$ GeV are lower by a factor of 2--3 compared to those in an inclusive search (see \Fig{fig:Prf2aOSisoPlots}), this cut allows us to strengthen the expected limit on $\sigma(pp\to V+X) \, \BR (V\to\mu^+\mu^-)$ for model M1 by $\gtrsim 2$ over a substantial portion of the kinematically allowed range.%
\footnote{Note that, in this model and within the mass range of interest, the efficiencies are weak functions of the $\ptmm$ cut.}

%

The largest improvement comes in the range $m_{\mu\mu} \in [35,55] \GeV$, where our expected bounds from the isolated sample for $\ptmm>25 \GeV$ are in the range of 35--45~fb.
For $m_V=m_A = 40 \GeV$, we estimate  $A_V(0)=54\%$, $A_V(25)=$ 47\%, $\epsilon^V_{\rm tr}\sim 85\%$, and $\epsilon^V_{\rm iso}\sim  85\%$.
(We will discuss these efficiencies further in future work; the isolation efficiency $\epsilon^V_{\rm iso}$ is smaller than in \Tab{tab:AEff} because the muons are softer and the Higgs process is accompanied by more initial state radiation.)
Using the observed bound from \Fig{fig:Prf2aOSisoPlots}, we obtain a limit for $m_V=m_A = 40 \GeV$ of  
\begin{align}
  \label{eq:Brh2mumuX}
  \BR(h\to Va){\rm} \, \BR(V\to\mu\mu) \lesssim 7 \times 10^{-3},
\end{align}
where we have conservatively taken the uncertainty on the 7 TeV total Higgs cross section to be $30\%$ with a flat prior.
Because of the high relative signal acceptance, $A_V(25)/A_V(0)$ of 85\%, this limit is more than a factor of 2.5 lower than what is expected when no $\ptmm$ cut is applied.
A simple scaling of our model-independent result suggests that limits of better than $\sim 10^{-3}$ could be expected from LHC Run II data, even after a penalty from higher trigger thresholds. 

Of course, a search targeted specifically for this model could obtain even stronger limits through an $m_V$- and $m_a$-dependent $\ptmm$ cut and by adding the $V\to e^+e^-$ channel.
In this context, it is interesting to consider some other models to which our limit applies and which have been constrained by existing analyses.
Both CMS \cite{Khachatryan:2017mnf} and ATLAS \cite{Aaboud:2018esj} have searched for $h\to a a\to (b\bar b) (\mu\mu)$, whose signature is identical to ours if $a\to b\bar b$ and $m_V=m_a$.  Both analyses required two $b$-tagged jets, and constrain the jets and muons to reconstruct a Higgs; ATLAS further requires that the invariant mass of the jets be similar to that of the muons. 
Using 19.7 fb$^{-1}$ of 8 TeV data, CMS obtained a limit (for $m_a = 40$ GeV) of $4\times 10^{-4}$, also achieved by ATLAS with 36.1 fb$^{-1}$ of  13 TeV data.
The order of magnitude improvement compared to \Eq{eq:Brh2mumuX} is not surprising considering the higher energy and integrated luminosity, along with the optimized targeting of a particular model which greatly reduces background.  
Of course, our limit continues to apply with little change even if $m_a\ll m_V$, or to variants of model M1 where the $a$ does not decay to $b\bar b$, situations to which the ATLAS and CMS limits do not generally apply.  
This illustrates the complementarity of targeted and model-independent search strategies, and the importance of each.

If $m_S>m_h$, then the \nameofsearch strategy yields a higher relative acceptance, and the $p_T$ cut can be raised.
As an example, we show in \Fig{fig:M1_125ratio25_200ratio60} (right) the relative acceptance $A_V(60)/A_V(0)$ of the dimuon $\ptmm>60$ GeV cut, for $m_S\sim 200$ GeV.  
With this $\ptmm$ cut, expected limits on $\sigma(pp\to V+X) \, \BR (V\to\mu^+\mu^-)$  can improve by as much as a factor of 5 relative to an inclusive search.

For model M2, either the isolated or prompt samples could yield the stronger limit, depending on the precise mass hierarchy.
Specifically, in the regime $m_S\gg m_{\chi_2}$, the $\chi_2$ is boosted, so the $V$ and $\chi_1$ produced in its decay are both boosted and collimated, as are their decay products.
Therefore, the muon isolation efficiency for the $V$ signal will be degraded, and the prompt sample may give better limits in this regime.
We relegate further details about M2 to future work.
Here we simply note that, according to our \textsc{Pythia} 8 simulation, both $A_V(25)/A(0)$ for $m_S\sim 125$ GeV and $A_V(60)/A(0)$ for $m_S\sim 200$ GeV are much higher than 50\% in much of the kinematic range, again implying that a \nameofsearch search can significantly outperform an inclusive search.

Beyond dimuon resonance searches, there are other LHC analyses that could be sensitive to models such as M1 and M2.
If the $S$ is the Higgs boson or is produced by mixing with the Higgs, then $WS$ and $ZS$ production rates are not negligible.  
In such cases, the \nameofsearch search described here should be compared not only with an inclusive search of the dimuon spectrum but also with multilepton searches.  
At the same integrated luminosity, the multilepton signal from $S$ is two orders of magnitude smaller than the total $S$ cross section, but in certain kinematic regimes it has small backgrounds.
The sensitivity of the two classes of searches may depend on the model and its parameters, and on the integrated luminosity, as well as on the specific design of the multilepton search, whose efficiencies and acceptance for low-$p_T$ leptons must be carefully accounted for.  
We have not attempted to make a detailed comparison, but for the model and parameters corresponding to our limit $h\to V+X$ in M1, \Eq{eq:Brh2mumuX}, fewer than four multilepton events arise for $\cL=2.11$\,fb\inv, before accounting for efficiencies and acceptance.  
Even with the full Run I data set, losses due to efficiencies and acceptance suggest that a limit from multilepton searches will not dramatically improve on \Eq{eq:Brh2mumuX}. 
Run II multilepton searches at ATLAS and CMS~(such as \Refs{Khachatryan:2017qgo,CMS-PAS-EXO-18-005,Sirunyan:2017qkz}) presumably could put stronger limits than we could achieve using CMS11a, but it is not obvious how they would compare with our method applied to the full Run II data set; a detailed study would be required.

However, if $S$ is produced not by mixing with the Higgs but through a separate coupling to gluons, then the $WS$ and $ZS$ processes are absent, eliminating the multilepton signal.
And if the muons are often non-isolated, the multilepton search loses its sensitivity.
In such cases, our \nameofsearch dimuon search competes only with inclusive dimuon searches, and often performs better, as we have already seen.
It seems likely that this is true for many other models in which a high-$p_T$ dilepton resonance is the dominant observable effect.
For such models, any limits obtained from the results presented here may potentially improve upon existing public limits, though a complete study of the Run II literature would be needed to confirm this. 

Most importantly, when applied to the Run II data set, the  \nameofsearch  search strategy should give bounds that are several times smaller than a Run II inclusive search, and up to an order of magnitude below those presented here.
We therefore view the discovery potential of this strategy as noteworthy. 

\subsection{Production of $V$ via Kinetic Mixing:  The Dark Photon Scenario}

By contrast, our \nameofsearch search strategy is not effective, and indeed counterproductive, for the popular benchmark dimuon resonance scenario known as the minimal dark photon model~\cite{Okun:1982xi,Galison:1983pa,Holdom:1985ag,Pospelov:2007mp,ArkaniHamed:2008qn,Bjorken:2009mm}.
Here, $V$ is predominantly produced via kinetic mixing with the photon/$Z$ of strength $\epsilon$, and the $p_T$ distribution of the signal is the same as for the DY background.
Consequently, any cut on $\ptmm$ reduces sensitivity to $\epsilon$, because it removes signal without changing $S/B$.
(As discussed in \Ref{Hoenig:2014dsa}, imposing a cut on $\ptmm$ is still useful to avoid the turn-on behavior of the dimuon trigger.)

Nevertheless, our inclusive search in the isolated sample for $m_{\mu\mu}>35$ GeV can be  compared to previous results.
At present, LHCb has the best LHC limits in the 10.6--70~GeV mass range~\cite{Ilten:2016tkc,Aaij:2017rft}, though BaBar is more sensitive below 10~GeV~\cite{Lees:2014xha} and future ATLAS/CMS searches are expected to be more sensitive above 40~GeV~\cite{Curtin:2014cca}.
(For a recent study of different dark photon and vector resonance bounds, see~\Refs{Ilten:2018crw,Bauer:2018onh}.)
The LHCb data sample has lower integrated luminosity (1.6 fb\inv) and narrower $\eta$ acceptance than the CMS11a sample, but the higher production rate at 13~TeV more than compensates.
Thus, in the region above 35 GeV, our limits on $\epsilon$ from the $\ptmm>0$ subsample should be comparable to but slightly weaker than those of LHCb~\cite{Aaij:2017rft}.
Following the analysis of \Ref{Ilten:2016tkc}, we obtain an estimated limit of $\epsilon^2 \lesssim1.3 \times 10^{-5}$ at $m_V=50$ GeV, which confirms this expectation.%
\footnote{A less stringent limit was estimated in \Ref{Hoenig:2014dsa} due to a more conservative treatment of the dimuon mass resolution.}
At lower $m_V$, where the trigger effectively already applies a $\ptmm$ cut, our limits on $\epsilon$ are further weakened.

\section{Discussion}
\label{sec:discussion}

Using $2.11$ fb\inv\ of CMS Open Data from 2011, we performed a model-independent \nameofsearch search for a new particle $V$ decaying to dimuons.
We showed how exploiting moderately boosted kinematics can give significantly lower bounds on a product of physics and detector quantities, because a simple $p_T$ cut on the dimuon system sharply reduces QCD and DY backgrounds.
As long as $V$ is typically produced in the decay of a heavier particle, this type of cut often preserves signal acceptance, and so our results will lead to improved limits on a wide class of models.
Our results indicate that limits in some classes of signal models can improve by up to a factor of 9 relative to those from an inclusive dimuon search at the same luminosity.
Still greater improvements could be achieved in some models by using even stronger $\ptmm$ cuts. 
A similar strategy would be relevant for diphoton resonances from a particle produced mainly in decays; see \Refs{Strassler:2006im,Chang:2006bw,Juknevich:2009ji}.

We argued that there exist reasonable and simple models for which a \nameofsearch  search would set better limits than any other search strategy implemented to date.
Though we only studied the dimuon final state, a combination with dielectrons would further improve the limit on many models.
With the much  larger integrated luminosity collected during Run II and the higher signal cross sections at 13 TeV (partially counter-balanced by higher trigger thresholds), we estimate that our bounds could shrink by an order of magnitude.
Thus  in LHC Run II data, the \nameofsearch search strategy would have considerable discovery potential for a diverse collection of theoretical models, over a wide range of resonance masses.

We have also emphasized the importance of searching both with and without imposing an isolation cut on the leptons.
Backgrounds increase by a factor of order 2 when the isolation cut is dropped and replaced with a stringent IP cut.
On the other hand, in models where the leptons are embedded in a cluster of particles produced in a hidden sector \cite{Strassler:2006im,Han:2007ae,Baumgart:2009tn}, the dimuon isolation efficiency may easily be smaller than order $1/\sqrt{2}$, such that the prompt sample provides {\it more} sensitivity than the isolated sample.

Finally, we have illustrated for the first time that open collider data has the potential to assist the BSM search program at the LHC.
In carrying out a search whose results, while limited, do probe new ground, we hope we have demonstrated two things.
First, open data can be used to study questions which are outside the mainstream search program, and thus explore new territory.
Second, when important backgrounds are challenging for theorists to simulate reliably, open data can provide those backgrounds directly, making phenomenological studies or prototype analyses far more accurate.
As an example, our prompt sample has large QCD backgrounds, and we could not have selected our IP cuts with confidence without the explicit knowledge of the backgrounds obtained from the CMS Open Data.
In our view, although searches using current open data are unlikely to uncover BSM phenomena on their own, they can help demonstrate the value of certain search strategies and justify the application of those strategies by the experimental collaborations on much larger data sets.

\acknowledgments{
We thank CERN, the CMS collaboration, and the CMS Data Preservation and Open Access (DPOA) team for making research-grade collider data available to the public. 
We thank R.~Leane, R.~Mastandrea, and especially R.~D'Agnolo for assistance at certain points in this work.
We thank E.~Carrera, K.~Lassila-Perini, and the CMS DPOA team for help interpreting the muon information in the CMS Open Data.
We are grateful for conversations with K.~Cranmer,  A.~Geiser, S.~Gori, B.~Nachman, S.~Somalwar, and especially  P.~Harris, S.~Rappoccio, and M.~Williams whose comments on a preliminary draft contributed to significant improvements in our methods.  
C.C.~is supported by the Office of High Energy Physics of the U.S. Department of Energy (DOE) under grant DE-SC0013607.
Y.S.~thanks the Aspen Center for Physics for support.
M.J.S.~thanks the Department of Physics at Harvard University for hospitality.
J.T.~is supported by the DOE under grant DE-SC0012567 and by the Simons Foundation through a Simons Fellowship in Theoretical Physics.
W.X.~is supported by the DOE under grants DE-SC0012567 and DE-SC0013999 and by the European Research Council grant NEO-NAT.  
}

\bibliographystyle{utphys}
\bibliography{mumu_bib}

\end{document}